\documentclass[english,3p, times, a4paper, 10pt, oneside, final]{elsarticle}
\usepackage[T1]{fontenc}
\usepackage[latin9]{inputenc}
\usepackage{xcolor}
\usepackage{textcomp}
\usepackage{mathrsfs}
\usepackage{amsmath}
\usepackage{amssymb}
\usepackage{graphicx}
\usepackage{wasysym}
\usepackage{esint}
\PassOptionsToPackage{normalem}{ulem}
\usepackage{ulem}
\usepackage[english]{babel}
\usepackage[absolute]{textpos}

\makeatletter
\usepackage{lineno,hyperref}
\usepackage{textcomp}
\usepackage{caption}
\graphicspath{{fig/}}

\hypersetup{
	pdfauthor={Markus Kantner},
    pdftitle={Generalized Scharfetter--Gummel schemes for electro-thermal transport in degenerate semiconductors using the Kelvin formula for the Seebeck coefficient},
    pdfkeywords={finite volume Scharfetter--Gummel method, semiconductor device simulation,
electro-thermal transport, non-isothermal drift-diffusion system,
degenerate semiconductors (Fermi--Dirac statistics), Seebeck coefficient, thermopower, energy model},
    colorlinks=true,
}

\makeatother

\begin{document}
\journal{Journal of Computational Physics}

\biboptions{longnamesfirst, sort&compress}

\newcommand{\antiderivative}{\mathscr{G}}
\newcommand{\ee}{\mathrm{e}}
\newcommand{\heatCapacity}{c_{V}}
\newcommand{\speciesIndex}{\lambda}

\begin{frontmatter}{}

\title{Generalized Scharfetter--Gummel schemes for electro-thermal transport
in degenerate semiconductors using the Kelvin formula for the Seebeck
coefficient}

\begin{textblock*}{200mm}(17.5mm,266mm)
\footnotesize \copyright~2019. Licensed under the Creative Commons \href{https://creativecommons.org/licenses/by-nc-nd/4.0/legalcode}{CC-BY-NC-ND 4.0}.\normalsize
\end{textblock*}

\author{Markus Kantner}

\address{Weierstrass Institute for Applied Analysis and Stochastics (WIAS),\\
Mohrenstr. 39, 10117 Berlin, Germany}

\ead{kantner@wias-berlin.de}

\begin{abstract}
Many challenges faced in today's semiconductor devices are related
to self-heating phenomena. The optimization of device
designs can be assisted by numerical simulations using the non-isothermal
drift-diffusion system, where the magnitude of the thermoelectric
cross effects is controlled by the Seebeck coefficient. We show
that the model equations take a remarkably simple form when assuming
the so-called Kelvin formula for the Seebeck coefficient. The corresponding
heat generation rate involves exactly the three classically known
self-heating effects, namely Joule, recombination and Thomson--Peltier
heating, without any further (transient) contributions. Moreover,
the thermal driving force in the electrical current density expressions
can be entirely absorbed in the diffusion coefficient
via a generalized Einstein relation. The efficient numerical simulation
relies on an accurate and robust discretization technique for the
fluxes (finite volume Scharfetter--Gummel method), which allows to
cope with the typically stiff solutions of the semiconductor device
equations. We derive two non-isothermal generalizations of
the Scharfetter--Gummel scheme for degenerate semiconductors (Fermi--Dirac
statistics) obeying the Kelvin formula. The approaches differ in
the treatment of degeneration effects: The first is based on an approximation
of the discrete generalized Einstein relation implying a specifically
modified thermal voltage, whereas the second scheme follows the conventionally
used approach employing a modified electric field. We present a detailed
analysis and comparison of both schemes, indicating a superior performance
of the modified thermal voltage scheme.
\end{abstract}


\end{frontmatter}{}

\section{Introduction}

Thermal effects cause many challenges in a broad variety of semiconductor
devices. Thermal instabilities limit the safe-operating area of high
power devices and modules in electrical energy technology \citep{Lutz2011,Schulze2012},
electro-thermal feedback loops lead to catastrophic snapback phenomena
in organic light-emitting diodes \citep{Fischer2014,Fischer2018}
and self-heating effects decisively limit the achievable output power
of semiconductor lasers \citep{Osinski1994,Piprek2002,Streiff2005,Wenzel2010}.
The numerical simulation of semiconductor devices
showing strong self-heating and thermoelectric effects requires a
thermodynamically consistent modeling approach, that describes the
coupled charge carrier and heat transport processes. In the context
of semiconductor device simulation, the non-isothermal drift-diffusion
system\emph{ }\citep{Wachutka1990,Lindefelt1994,Brand1995,Parrott1996,Albinus2002,Bandelow2005}
has become the standard model for the self-consistent description
of electro-thermal transport phenomena. This is a system of four partial
differential equations, which couples the semiconductor device equations
\citep{Selberherr1984,Markowich1986} to a (lattice) heat flow equation
for the temperature distribution in the device. On the step from the
\emph{isothermal} to the \emph{non-isothermal} drift-diffusion system,
additional thermoelectric transport coefficients must be included
in the theory. The magnitude of the thermoelectric cross-effects is
governed by the Seebeck coefficient (also\emph{ thermopower}), which
quantifies the thermoelectric voltage induced by a temperature gradient
(\emph{Seebeck effect}) \citep{Goldsmid2010,Goupil2011}. The reciprocal
phenomenon of the Seebeck effect is the \emph{Peltier effect}, which
describes the current-induced heating or cooling at material junctions.
As a consequence of Onsager's reciprocal relations, the Seebeck and
Peltier coefficients are not independent such that only the Seebeck
coefficient must be specified \citep{Onsager1931}. Over the decades,
several definitions have been proposed for the Seebeck coefficient
\citep{Kubo1957b,Cutler1969,Fritzsche1971,Chaikin1976}; recent publications
list at least five coexisting different (approximate) formulas \citep{Shastry2013,Freeman2014}.
In the context of semiconductor device simulation, the Seebeck coefficients
are typically derived from the Boltzmann transport equation in relaxation
time approximation \citep{VanVliet1976,Marshak1984,Lundstrom2000}
or defined according to the adage of the Seebeck coefficient being
the ``(specific) entropy per carrier'' \citep{Albinus2002,Bandelow2005,Goupil2011,Wenzel2017}.
These approaches are often focused on non-degenerate semiconductors,
where the carriers follow the classical Maxwell--Boltzmann statistics.
This approximation breaks down in heavily doped semiconductors, where
the electron-hole plasma becomes degenerate and Fermi--Dirac statistics
must be considered to properly take into account the Pauli exclusion
principle. Degeneration effects are important in many semiconductor
devices such as semiconductor lasers, light emitting diodes or transistors.
Moreover, heavily doped semiconductors are considered as ``good''
thermoelectric materials, i.\,e., materials with high thermoelectric
figure of merit \citep{Goldsmid2010,Goupil2011}, for thermoelectric
generators, which can generate electricity from waste heat \citep{Snyder2008,Bennett2017}.

In this paper, we will consider an alternative model for the Seebeck
coefficient, which is the so-called \emph{Kelvin formula for the thermopower}
\citep{Peterson2010}. The Kelvin formula recently gained interest
in theoretical condensed matter physics and has been shown to yield
a good approximation of the Seebeck coefficient for many materials
(including semiconductors, metals and high temperature superconductors)
at reasonably high temperatures \citep{Shastry2008,Silk2009,Peterson2010,Garg2011,Arsenault2013,Deng2013,Zlatic2014,Kokalj2015,Hejtmanek2015,Terasaki2016,Mravlje2016}.
The Kelvin formula relates the Seebeck coefficient to the derivative
of the entropy density with respect to the carrier density and therefore
involves only equilibrium properties of the electron-hole plasma,
where degeneration effects are easily included. To our knowledge,
the Kelvin formula has not been considered in the context of semiconductor
device simulation so far. In Sec.~\ref{sec: The energy drift diffusion model and the Kelvin formula for the Seebeck coefficent},
we show that the Kelvin formula yields a remarkably simple form of
the non-isothermal drift-diffusion system, which shows two exceptional
features:
\begin{enumerate}
\item The heat generation rate involves exactly the three classically known
self-heating effects (Joule, Thomson--Peltier and recombination heating)
without any further (transient) contributions.
\item The thermal driving force in the current density expressions can
be entirely absorbed in a (nonlinear) diffusion coefficient via a
generalized Einstein relation. Hence, the $\nabla T$ term is eliminated
in the drift-diffusion form.
\end{enumerate}
The second part of this paper (Sec.~\ref{sec: Non-isothermal generalization of the Scharfetter=002013Gummel scheme for degenerate semiconductors})
deals with the discretization of the electrical current density expressions,
which are required in (non-isothermal) semiconductor device simulation
tools. The robust and accurate discretization of the drift-diffusion
fluxes in semiconductors with exponentially varying carrier densities
is a non-trivial problem, that requires a special purpose discretization
technique. The problem has been solved by Scharfetter and Gummel
for the case of non-degenerate semiconductors under isothermal conditions
\citep{Scharfetter1969}. Since then, several adaptations of the method
have been developed to account for more general situations (non-isothermal
conditions \citep{Tang1984,McAndrew1985,Rudan1986,Forghieri1988,Chen1991,Souissi1991,TenThijeBoonkkamp1993,Smith1993},
degeneration effects \citep{Yu1985,Gajewski1993,Bessemoulin-Chatard2012,Koprucki2013,Koprucki2015,Fuhrmann2015}).
The Kelvin formula for the Seebeck coefficients allows for a straightforward
generalization of the Scharfetter--Gummel approach to the non-isothermal
case. We take up two different approaches to incorporate degeneration
effects into the non-isothermal Scharfetter--Gummel formula and give
an extensive numerical and analytical comparison of both methods.
This includes an investigation of limiting cases and structure preserving
properties of the discrete formulas (Sec.~\ref{Sec: Limiting cases and structure preserving properties}),
a comparison with the numerically exact solution of the underlying
two-point boundary value problem (Sec.~\ref{sec: Comparison with numerically exact solution})
and a comparison of analytical error bounds (Sec.~\ref{sec: Analytical error estimate}).
Finally, in Sec.~\ref{sec: benchmark simulation}, we present a numerical
convergence analysis of both schemes based on numerical simulations
of a one-dimensional p-n-diode.

\section{The non-isothermal drift-diffusion system using the Kelvin formula
for the Seebeck coefficient \label{sec: The energy drift diffusion model and the Kelvin formula for the Seebeck coefficent}}

\begin{figure}
\includegraphics[width=1\textwidth]{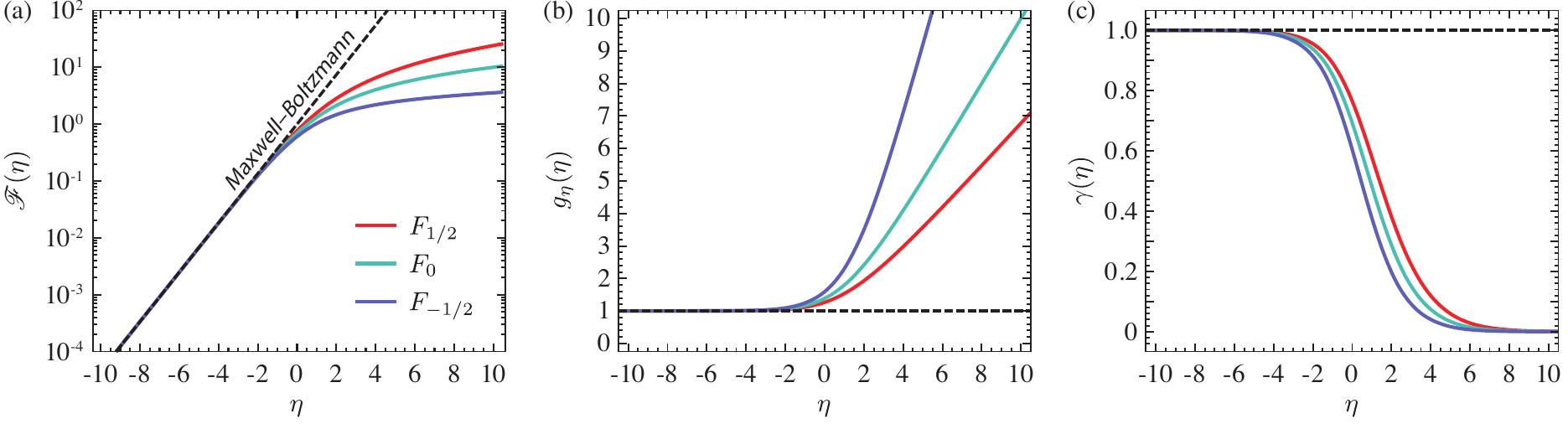}

\caption{(a)~Fermi--Dirac integrals (\ref{eq: Fermi-Dirac integral}) of
order $\nu=1/2$, $0$ and $-1/2$ as functions of the reduced Fermi
energy $\eta$. For $\eta\ll-1$ the Fermi--Dirac integrals approach
the Maxwell--Boltzmann distribution $\mathscr{F}\left(\eta\right)=\exp{\left(\eta\right)}$ (non-degenerate limit).
(b)~Plot of the degeneracy factor (\ref{eq: degeneracy factor - eta})
(or diffusion enhancement factor) for the Fermi--Dirac integrals
in (a). For $\eta\ll-1$ the degeneracy factor approaches $1$ (linear
diffusion). (c)~Correction factor (\ref{eq: correction factor-1-1})
that quantifies the deviation of the Fermi--Dirac integrals from
the exponential function. The non-degenerate limit corresponds to
$\gamma\left(\eta\right)\equiv1$.}

\label{fig: distribution function and degeneracy factor}
\end{figure}

In this section we briefly review the non-isothermal drift-diffusion
system, which provides a self-consistent description of the coupled
electro-thermal transport processes in semiconductor devices. The
model has been extensively studied by several authors from the perspective
of physical kinetics or phenomenological non-equilibrium thermodynamics
\citep{Wachutka1990,Lindefelt1994,Brand1995,Parrott1996,Albinus2002,Bandelow2005}.
The model equations read:
\begin{align}
-\nabla\cdot\varepsilon\nabla\phi & =q\left(C+p-n\right),\label{eq: Poisson equation}\\
\partial_{t}n-\frac{1}{q}\nabla\cdot\mathbf{j}_{n} & =-R,\label{eq: electron transport equation}\\
\partial_{t}p+\frac{1}{q}\nabla\cdot\mathbf{j}_{p} & =-R,\label{eq: hole transport equation}\\
\heatCapacity\partial_{t}T-\nabla\cdot\kappa\nabla T & =H.\label{eq: heat equation}
\end{align}
Poisson's equation~(\ref{eq: Poisson equation}) describes the electrostatic
field generated by the electrical charge density $\rho=q\left(C+p-n\right)$.
Here, $\phi$ is the electrostatic potential, $n$ and $p$ are the
densities of electrons and holes, respectively, $C$ is the built-in
doping profile, $q$ is the elementary charge and $\varepsilon$ is
the (absolute) dielectric constant of the material. The transport
and recombination dynamics of the electron-hole plasma are modeled
by the continuity equations (\ref{eq: electron transport equation})--(\ref{eq: hole transport equation}),
where $\mathbf{j}_{n/p}$ are the electrical current densities and
$R$ is the (net-)recombination rate. The latter includes several
radiative and non-radiative recombination processes (Shockley--Read--Hall
recombination, Auger recombination, spontaneous emission etc.) \citep{Selberherr1984}.
The carrier densities $n$, $p$ are connected with the electrostatic
potential $\phi$ via the state equations
\begin{align}
n & =N_{c}\left(T\right)\mathscr{F}\left(\frac{\mu_{c}+q\phi-E_{c}\left(T\right)}{k_{B}T}\right), & p & =N_{v}\left(T\right)\mathscr{F}\left(\frac{E_{v}\left(T\right)-q\phi-\mu_{v}}{k_{B}T}\right),\label{eq: carrier density state equations}
\end{align}
where $k_{B}$ is Boltzmann's constant, $T$ is the absolute temperature,
$N_{c/v}$ is the effective
density of states and $E_{c/v}$ is the reference energy level (typically
the band edge energy) of the conduction band or valence band, respectively.
The function $\mathscr{F}$ describes the occupation of the electronic
states under \emph{quasi-equilibrium} conditions, which is controlled
by the quasi-Fermi energies $\mu_{c/v}$ of the respective bands.
The quasi-Fermi energies are connected with the quasi-Fermi potentials
$\varphi_{n/p}$ via
\begin{align}
\mu_{c} & =-q\varphi_{n}, & \mu_{v} & =-q\varphi_{p}.\label{eq: quasi-Fermi potentials-1}
\end{align}
In non-degenerate semiconductors (Maxwell--Boltzmann statistics),
$\mathscr{F}$ is the exponential function $\mathscr{F}\left(\eta\right)=\exp{\left(\eta\right)}$.
Taking the degeneration of the electron-hole plasma due to Pauli-blocking
into account (Fermi--Dirac statistics), $\mathscr{F}$ is typically
given by the Fermi--Dirac integral
\begin{equation}
\mathscr{F}\left(\eta\right)=F_{\nu}\left(\eta\right)=\frac{1}{\Gamma\left(\nu+1\right)}\int_{0}^{\infty}\mathrm{d}\xi\,\frac{\xi^{\nu}}{\exp{\left(\xi-\eta\right)}+1},\label{eq: Fermi-Dirac integral}
\end{equation}
where the index $\nu$ depends on the dimensionality of the structure.
Isotropic, bulk materials with parabolic energy bands are described
by $\nu=1/2$; for two-dimensional materials (quantum wells) the
index $\nu=0$ applies. See Fig.~\ref{fig: distribution function and degeneracy factor}\,(a)
for a plot of the Fermi--Dirac integrals for different $\nu$ as
a function of the reduced Fermi energy.
The function $\mathscr{F}$ may also include non-parabolicity effects, see \ref{sec: non-parabolic energy dispersion}.
In the case of organic semiconductors,
$\mathscr{F}$ is often taken as the Gauss--Fermi integral \citep{Mensfoort2008,Paasch2010}
or a hypergeometric function \citep{Vissenberg1998,Seki2013}.

The heat transport equation~(\ref{eq: heat equation}) describes
the spatio-temporal dynamics of the temperature distribution in the
device. Here, $\heatCapacity$ is
the (volumetric) heat capacity, $\kappa$ is the thermal conductivity
and $H$ is the heat generation rate. The non-isothermal drift-diffusion
model assumes a local thermal equilibrium between the lattice and
the carriers, i.\,e., $T=T_{L}=T_{n}=T_{p}$. The system (\ref{eq: Poisson equation})--(\ref{eq: heat equation})
must be supplemented with initial conditions and boundary conditions (i.e., for electrical contacts,
semiconductor-insulator interfaces, heat sinks etc). We refer to Refs.~\citep{Selberherr1984,Palankovski2004}
for a survey on commonly used boundary condition models.

The electrical current densities are driven by the gradients
of the quasi-Fermi potentials and the temperature
\begin{align}
\mathbf{j}_{n} & =-\sigma_{n}\left(\nabla\varphi_{n}+P_{n}\nabla T\right), & \mathbf{j}_{p} & =-\sigma_{p}\left(\nabla\varphi_{p}+P_{p}\nabla T\right),\label{eq: current densities-2}
\end{align}
where $\sigma_{n}=qM_{n}n$ and $\sigma_{p}=qM_{p}p$ are the electrical
conductivities (with carrier mobilities $M_{n/p}$) and $P_{n/p}$
are the Seebeck coefficients. Finally, we consider a (net-)recombination rate of the form \citep{Kantner2018c}
\begin{equation} 
R = R\left(\phi,\varphi_{n},\varphi_{p},T\right) = \left(1-\exp{\left(-\frac{\mu_{c}-\mu_{v}}{k_{B}T}\right)}\right)\sum_{\alpha}r_{\alpha}\left(\phi,\varphi_{n},\varphi_{p},T\right), \label{eq: recombination rate}
\end{equation} 
which combines several radiative and non-radiative recombination processes labeled by $\alpha$ (e.g., Shockley--Read--Hall recombination, spontaneous emission, Auger recombination etc.). The functions $r_{\alpha}=r_{\alpha}\left(\phi,\varphi_{n},\varphi_{p},T\right)\geq 0$ are inherently non-negative and specific for the respective processes. We refer to Refs.~\cite{Selberherr1984,Palankovski2004} for commonly considered recombination rate models.

\subsection{Kelvin formula for the Seebeck coefficient} \label{sec: Kelvin formula for the Seebeck coefficient}

In this paper, we consider the so-called \emph{Kelvin formula}
for the Seebeck coefficient \citep{Peterson2010}
\begin{align}
P_{n} & =-\frac{1}{q}\frac{\partial s\left(n,p,T\right)}{\partial n}, & P_{p} & =+\frac{1}{q}\frac{\partial s\left(n,p,T\right)}{\partial p},\label{eq: Kelvin formula}
\end{align}
which relates the thermoelectric powers to the derivatives of the
entropy density $s=s(n,p,T)$ with respect to the carrier densities.
The expression for the entropy density is easily derived from  the
free energy density $f\left(n,p,T\right)$ of the system, which is
a proper thermodynamic potential if the set of unknowns is chosen
as $\left(n,p,T\right)$ (``natural variables'') . The expressions
for the quasi-Fermi energies and the entropy density then follow as
\begin{align}
\frac{\partial f\left(n,p,T\right)}{\partial n} & =+\mu_{c}\left(n,p,T\right), & \frac{\partial f\left(n,p,T\right)}{\partial p} & =-\mu_{v}\left(n,p,T\right), & \frac{\partial f\left(n,p,T\right)}{\partial T} & =-s\left(n,p,T\right).\label{eq: conjugate fields-1}
\end{align}
Taking the second derivatives, this yields the Maxwell relations
\begin{align}
\frac{\partial\mu_{c}\left(n,p,T\right)}{\partial T} & =-\frac{\partial s\left(n,p,T\right)}{\partial n}, & \frac{\partial\mu_{v}\left(n,p,T\right)}{\partial T} & =+\frac{\partial s\left(n,p,T\right)}{\partial p},\label{eq: Maxwell relations-1}
\end{align}
which allow for an alternative representation of Eq.~(\ref{eq: Kelvin formula}).
The free energy density includes contributions from the quasi-free
electron-hole plasma (ideal Fermi gas), the lattice vibrations (ideal
Bose gas) and the electrostatic (Coulomb) interaction energy. Throughout
this paper, we assume a free energy density of the form \citep{Albinus2002}
\begin{equation}
f\left(n,p,T\right)=f_{\text{e--h}}\left(n,p,T\right)+f_{L}\left(T\right)+f_{\text{Coul}}\left(p-n\right).\label{eq: free energy density-1}
\end{equation}
The free energy density of the (non-interacting) electron-hole plasma
reads \citep{Albinus2002,Kantner2018c}
\begin{align}
\begin{aligned}f_{\text{e--h}}\left(n,p,T\right) & =k_{B}T\mathscr{F}^{-1}\left(\frac{n}{N_{c}\left(T\right)}\right)n-k_{B}TN_{c}\left(T\right)\antiderivative\left(\mathscr{F}^{-1}\left(\frac{n}{N_{c}\left(T\right)}\right)\right)+E_{c}\left(T\right)n\\
 & \hphantom{=}+k_{B}T\mathscr{F}^{-1}\left(\frac{p}{N_{v}\left(T\right)}\right)p-k_{B}TN_{v}\left(T\right)\antiderivative\left(\mathscr{F}^{-1}\left(\frac{p}{N_{v}\left(T\right)}\right)\right)-E_{v}\left(T\right)p,
\end{aligned}
\label{eq: free energy density electron-hole plasma-1-1}
\end{align}
where $\mathscr{F}^{-1}$ is the inverse of the function $\mathscr{F}$
in the state equations (\ref{eq: carrier density state equations})
and $\antiderivative$ denotes its antiderivative: $\antiderivative^{\prime}\left(\eta\right)=\mathscr{F}\left(\eta\right)$.
Note that Eq.~(\ref{eq: free energy density electron-hole plasma-1-1})
implies\begin{subequations}\label{eq: free energy derivatives}
\begin{align}
\frac{\partial f_{\text{e--h}}}{\partial n} & =k_{B}T\mathscr{F}^{-1}\left(\frac{n}{N_{c}\left(T\right)}\right)+E_{c}\left(T\right), & \frac{\partial f_{\text{e--h}}}{\partial p} & =k_{B}T\mathscr{F}^{-1}\left(\frac{p}{N_{v}\left(T\right)}\right)-E_{v}\left(T\right).\label{eq: non-interacting Fermi gas chemical potentials-1}
\end{align}
The lattice contribution $f_{L}\left(T\right)$ yields the dominant
contribution to the heat capacity $\heatCapacity$. It can be derived
from, e.\,g., the Debye model for the free phonon gas \citep{Czycholl2008}.
The Coulomb interaction energy $f_{\text{Coul}}$ must be modeled
such that the state equations~(\ref{eq: carrier density state equations})
follow consistently from solving the defining relations for the quasi-Fermi
energies~(\ref{eq: conjugate fields-1}) for the carrier densities.
In order to supplement the ``missing'' electrostatic contributions in
Eq.~(\ref{eq: non-interacting Fermi gas chemical potentials-1}),
we specify the derivatives of $f_{\text{Coul}}$ with respect to the
carrier densities:
\begin{align}
\frac{\partial f_{\text{Coul}}}{\partial n} & =-q\phi, & \frac{\partial f_{\text{Coul}}}{\partial p} & =+q\phi.\label{eq: electrostatic derivatives-1}
\end{align}
\end{subequations}We refer to Albinus et al. \citep{Albinus2002}
for a rigorous mathematical treatment of the Coulomb interaction
energy.

\begin{figure}
\includegraphics[width=1\textwidth]{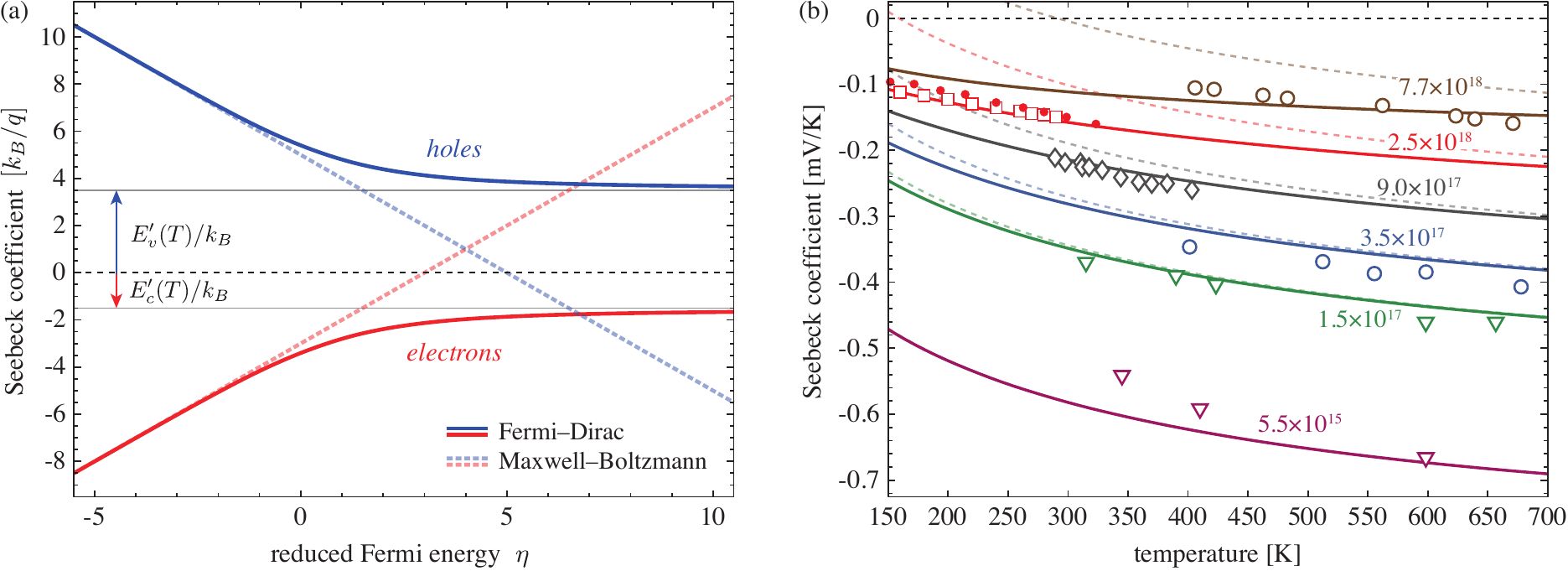}

\caption{(a)~Seebeck coefficient according to the Kelvin formula (\ref{eq: explicit Seebeck-1})
as a function of the reduced Fermi energy $\eta$ for power law type
effective density of states $N_{c/v}\propto T^{3/2}$ and $\mathscr{F}\left(\eta\right)=F_{1/2}\left(\eta\right)$
in units of $k_{B}/q$. The formula (\ref{eq: explicit Seebeck-1})
takes degeneration effects (Fermi--Dirac statistics, solid lines)
of the electron-hole plasma into account, which causes a deviation
from the non-degenerate result (Maxwell--Boltzmann statistics, dashed
lines) at $\eta\gtrsim-1$. The temperature dependency of the band
gap energy yields an offset of $E_{c}^{\prime}\left(T\right)=\left(\chi+\frac{1}{2}\right)E_{g}^{\prime}\left(T\right)$
for electrons (red lines) and $E_{v}^{\prime}=\left(\chi-\frac{1}{2}\right)E_{g}^{\prime}\left(T\right)$
for holes (blue lines). The plot is for $\chi=-0.2$ and $k_{B}^{-1}E_{g}^{\prime}\left(T\right)=-5$.
(b)~Seebeck coefficient for n-type GaAs. Solid lines are computed
according to the Kelvin formula (\ref{eq: explicit Seebeck - electrons-1})
using Fermi--Dirac statistics, dashed lines indicate the corresponding
non-degenerate limit. The respective ionized donor densities $C=N_{D}^{+}$
are given in the plot in units of $\text{cm}^{-3}$. The temperature-dependency
of the band gap energy $E_{g}\left(T\right)$ is modeled by the Varshni
model (\ref{eq: Varshni model}) with data from Ref.~\citep{Palankovski2004}
and the effective mass is $m_{c}^{\ast}\left(T\right)=\left(0.067-1.2\times10^{-5}\,\text{K}^{-1}\,T\right)m_{0}$
\citep{Palankovski2004}, where $m_{0}$ is the free electron mass.
The fitting parameter is set to $\chi=-0.2$. Experimental data: $\triangledown$
Carlson et al. \citep{Carlson1962}, $\Circle$ Amith et al. \citep{Amith1965},
$\lozenge$ Edmond et al. \citep{Edmond1956} (data from Ref.~\citep{Sutadhar1979}),
$\square$ Homm et al. \citep{Homm2008} and $\bullet$ Emel\textquoteright yanenko
et al. \citep{Emelyanenko1973} (data from Ref.~\citep{Sutadhar1979}).}

\label{fig: Seebeck coefficient-2}
\end{figure}

The Seebeck coefficients (\ref{eq: Kelvin formula}) are evaluated
using Eqs.~(\ref{eq: conjugate fields-1})--(\ref{eq: free energy derivatives}).
Since $f_{\text{Coul}}$ is independent of the temperature and $f_{L}$
does not depend on the carrier densities, the evaluation of Eq.~(\ref{eq: Kelvin formula})
requires only the Maxwell relations (\ref{eq: Maxwell relations-1})
and the derivatives of Eqs.~(\ref{eq: non-interacting Fermi gas chemical potentials-1})
with respect to the temperature. One obtains\begin{subequations}\label{eq: explicit Seebeck-1}
\begin{align}
P_{n}\left(n,T\right) & =-\frac{k_{B}}{q}\left(\frac{TN_{c}^{\prime}\left(T\right)}{N_{c}\left(T\right)}g\left(\frac{n}{N_{c}\left(T\right)}\right)-\mathscr{F}^{-1}\left(\frac{n}{N_{c}\left(T\right)}\right)-\frac{1}{k_{B}}E_{c}^{\prime}\left(T\right)\right),\label{eq: explicit Seebeck - electrons-1}\\
P_{p}\left(p,T\right) & =+\frac{k_{B}}{q}\left(\frac{TN_{v}^{\prime}\left(T\right)}{N_{v}\left(T\right)}g\left(\frac{p}{N_{v}\left(T\right)}\right)-\mathscr{F}^{-1}\left(\frac{p}{N_{v}\left(T\right)}\right)+\frac{1}{k_{B}}E_{v}^{\prime}\left(T\right)\right),\label{eq: explicit Seebeck - holes-1}
\end{align}
\end{subequations}where the prime denotes the derivatives
$N_{c/v}^{\prime}\left(T\right)=\partial_{T}N_{c/v}\left(T\right)$
and $E_{c/v}^{\prime}\left(T\right)=\partial_{T}E_{c/v}\left(T\right)$.
For power law type temperature dependency $N_{c/v}\left(T\right)\propto T^{\theta}$
(e.\,g., $\theta=3/2$), the factor in the first term reduces
to a constant $TN_{c/v}^{\prime}\left(T\right)/N_{c/v}\left(T\right)=\theta$.
For temperature-dependent effective masses, the term is more complicated.\emph{
}The function\begin{subequations}\label{eq: degeneracy factor - both expressions}
\begin{equation}
g\left(x\right)=x\,\frac{\mathrm{d}\mathscr{F}^{-1}\left(x\right)}{\mathrm{d}x}\label{eq: degeneracy factor-2-1}
\end{equation}
quantifies the degeneration of the Fermi gas. For non-degenerate
carrier statistics (Maxwell--Boltzmann statistics), Eq.~(\ref{eq: degeneracy factor-2-1})
reduces to exactly $g\equiv1$. For degenerate carrier statistics
one obtains $g>1$, which implies a nonlinear enhancement of the diffusion
current (see Sec.~\ref{sec: Drift-diffusion-current-densities}).
For later use, we also introduce the function
\begin{equation}
g_{\eta}\left(\eta\right)\equiv g\left(\mathscr{F}\left(\eta\right)\right)=\frac{\mathscr{F}\left(\eta\right)}{\mathscr{F}^{\prime}\left(\eta\right)},\label{eq: degeneracy factor - eta}
\end{equation}
\end{subequations}which is plotted in Fig.~\ref{fig: distribution function and degeneracy factor}\,(b).
The last terms in Eq.~(\ref{eq: explicit Seebeck-1}) describe the
contributions of the temperature dependency of the band edge energies
to the Seebeck coefficients. The two terms are not independent, as
they are required to satisfy $E_{g}^{\prime}\left(T\right)=E_{c}^{\prime}\left(T\right)-E_{v}^{\prime}\left(T\right)$,
where $E_{g}\left(T\right)$ is the energy band gap. A plot of the
Seebeck coefficients (\ref{eq: explicit Seebeck-1}) as functions
of the reduced Fermi energy $\eta$ is shown in Fig.~\ref{fig: Seebeck coefficient-2}\,(a)
for $\mathscr{F}\left(\eta\right)=F_{1/2}\left(\eta\right)$ and $N_{c,v}\propto T^{3/2}$.
The plot illustrates schematically the impact of the temperature derivatives
of the band edge energies and the role of degeneration effects.

In the following, several consequences of the Kelvin formula for the Seebeck coefficients are described, which are very appealing for numerical semiconductor device simulation as they greatly simplify the model equations.
Before going into details, we emphasize that the Kelvin formula is of course merely a convenient approximation and by no means exact.
More accurate and microscopically better justified approaches to calculate the Seebeck coefficient are based on
advanced kinetic models such as the semi-classical Boltzmann transport equation
beyond the relaxation time approximation (retaining the full form of the collision operator \cite{Ramu2010, Mascali2017}) 
or fully quantum mechanical methods \cite{Arsenault2013,Deng2013,Kokalj2015, Mravlje2016}.

\subsection{Comparison with experimental data}

Several empirical models for the temperature dependency of the band
gap energy have been proposed in the literature \citep{ODonnell1991},
including the commonly accepted Varshni model
\begin{equation}
E_{g}\left(T\right)=E_{g,0}-\frac{\alpha T^{2}}{\beta+T},\label{eq: Varshni model}
\end{equation}
where $E_{g,0}$, $\alpha$ and $\beta$ are material specific constants
\citep{Vurgaftman2001}. In order to specify $E_{c/v}^{\prime}\left(T\right)$
from Eq.~(\ref{eq: Varshni model}), we introduce a parameter $\chi$
such that $E_{c}^{\prime}\left(T\right)=\left(\chi+\frac{1}{2}\right)E_{g}^{\prime}\left(T\right)$
and $E_{v}^{\prime}\left(T\right)=\left(\chi-\frac{1}{2}\right)E_{g}^{\prime}\left(T\right)$.
In applications, $\chi$ can be used as a fitting parameter. It shall
be noted that the terms involving $E_{c/v}^{\prime}\left(T\right)$
in Eq.~(\ref{eq: explicit Seebeck-1}) are non-negligible and yield
a significant contribution to the Seebeck coefficients at elevated
temperatures. Indeed, some room temperature values of $k_{B}^{-1}E_{g}^{\prime}\left(300\,\text{K}\right)$
for important semiconductors are $-2.95$ (Si), $-4.47$ (Ge), $-5.32$
(GaAs) \citep{Palankovski2004}, which are on the same order of magnitude
as the first term $TN_{c/v}^{\prime}\left(T\right)/N_{c/v}\left(T\right)\approx1.5$
in Eq.~(\ref{eq: explicit Seebeck-1}).

In Fig.~\ref{fig: Seebeck coefficient-2}\,(b), the Kelvin formula
is plotted along with experimental data for n-GaAs. We observe a good
quantitative agreement of the formula (\ref{eq: explicit Seebeck - electrons-1})
with the experimental data in both the weak and the heavy doping regime
for temperatures above $150\,\text{K}$. At high carrier densities
(${N_{D}^{+}\geq9\times10^{17}\,\text{cm}^{-3}}$) the conduction
band electrons become degenerate (see the deviation of the solid from the
dashed lines), where the experimental values nicely follow the degenerate
formula (\ref{eq: explicit Seebeck - electrons-1}). See the caption
for details. At low temperatures ($T<150\,\text{K}$,
not shown), the Seebeck coefficient is increasingly dominated by the
phonon drag effect \citep{Homm2008}, which is not considered in
the present model.

\subsection{Heat generation rate \label{sec: Heat generation rate}}

A commonly accepted form of the self-consistent heat generation rate
$H$ was derived by Wachutka \citep{Wachutka1990}:
\begin{align}
\begin{aligned}H & =\frac{1}{\sigma_{n}}\left\Vert \mathbf{j}_{n}\right\Vert ^{2}+\frac{1}{\sigma_{p}}\left\Vert \mathbf{j}_{p}\right\Vert ^{2}-T\,\mathbf{j}_{n}\cdot\nabla P_{n}-T\,\mathbf{j}_{p}\cdot\nabla P_{p}+q\left(T\frac{\partial\varphi_{n}\left(n,p,T\right)}{\partial T}-\varphi_{n}-T\frac{\partial\varphi_{p}\left(n,p,T\right)}{\partial T}+\varphi_{p}\right)R\\
 & \phantom{=}-T\left(\frac{\partial\varphi_{p}\left(n,p,T\right)}{\partial T}+P_{p}\right)\nabla\cdot\mathbf{j}_{p}-T\left(\frac{\partial\varphi_{n}\left(n,p,T\right)}{\partial T}+P_{n}\right)\nabla\cdot\mathbf{j}_{n}.
\end{aligned}
\label{eq: Wachutka heat source}
\end{align}
Here we omit the radiation power density contribution from the original
work. The notation $\left\Vert \mathbf{x}\right\Vert =\left(\mathbf{x}\cdot\mathbf{x}\right)^{1/2}$
is the standard vector norm.
The derivation of Eq.~(\ref{eq: Wachutka heat source}) is based on the conservation of internal energy and does not involve any explicit assumptions on the Seebeck coefficient.
Using the Maxwell relations (\ref{eq: Maxwell relations-1}) and the transport
Eqs.~(\ref{eq: electron transport equation})--(\ref{eq: hole transport equation}),
we rewrite Eq.~(\ref{eq: Wachutka heat source}) as
\begin{align}
\begin{aligned}H & =\frac{1}{\sigma_{n}}\left\Vert \mathbf{j}_{n}\right\Vert ^{2}+\frac{1}{\sigma_{p}}\left\Vert \mathbf{j}_{p}\right\Vert ^{2}-T\,\mathbf{j}_{n}\cdot\nabla P_{n}-T\,\mathbf{j}_{p}\cdot\nabla P_{p}+q\left(\varphi_{p}-\varphi_{n}+\Pi_{p}-\Pi_{n}\right)R\\
 & \phantom{=}+qT\left(P_{p}-\frac{1}{q}\frac{\partial s\left(n,p,T\right)}{\partial p}\right)\partial_{t}p-qT\left(P_{n}+\frac{1}{q}\frac{\partial s\left(n,p,T\right)}{\partial n}\right)\partial_{t}n,
\end{aligned}
\label{eq: heat source-1}
\end{align}
where we introduced the Peltier coefficients $\Pi_{n}=TP_{n}$ and
$\Pi_{p}=TP_{p}$ (``second Kelvin relation''). Before we highlight the consequences of the Kelvin
formula for the Seebeck coefficients on $H$, we give a brief interpretation
of the individual terms in Eq.~(\ref{eq: heat source-1}).

The first two terms $H_{J,\speciesIndex}=\sigma_{\speciesIndex}^{-1}\left\Vert \mathbf{j}_{\speciesIndex}\right\Vert ^{2}$
(for $\speciesIndex\in\left\{ n,p\right\} $) describe Joule heating,
which is always non-negative and therefore never leads to cooling
of the device. The next two terms $H_{\text{T--P},\speciesIndex}=-T\,\mathbf{j}_{\speciesIndex}\cdot\nabla P_{\speciesIndex}$
(for $\speciesIndex\in\left\{ n,p\right\} $) describe the Thomson--Peltier
effect, which can either heat or cool the device depending on the
direction of the current flow. At constant temperature, this reduces
to the Peltier effect $H_{\text{T--P},\speciesIndex}\vert_{T=\text{const}.}=-\mathbf{j}_{\speciesIndex}\cdot\nabla\Pi_{\speciesIndex}$,
which is important at heterointerfaces and p-n junctions. At constant
carrier densities, one obtains the Thomson heat term $H_{\text{T--P},\speciesIndex}\vert_{n,p=\text{const}.}=-\mathcal{K}_{\speciesIndex}\,\mathbf{j}_{\speciesIndex}\cdot\nabla T$
with the Thomson coefficient $\mathcal{K}_{\speciesIndex}=T\frac{\partial P_{\speciesIndex}}{\partial T}=\frac{\partial\Pi_{\speciesIndex}}{\partial T}-P_{\speciesIndex}$
(for $\speciesIndex\in\left\{ n,p\right\} $). The Thomson--Peltier
effect combines both contributions. The recombination heat term $H_{R}=q(\varphi_{p}-\varphi_{n}+\Pi_{p}-\Pi_{n})R$
models the self-heating of the device due to recombination of electron-hole
pairs. The difference of the Peltier coefficients describes the
average excess energy of the carriers above the Fermi voltage. The
last line in Eq.~(\ref{eq: heat source-1}) is a purely transient
contribution, that has been discussed by several authors \citep{Wachutka1990,Lindefelt1994,Brand1995,Parrott1996,Freeman2014}.
In simulation practice, this term is often neglected, since estimations
show that it is negligible in comparison with the other self-heating
sources, see Refs.~\citep{Kells1993,Wolbert1994}.

We observe that the transient term vanishes exactly if we choose the
Kelvin formula (\ref{eq: Kelvin formula}) for the Seebeck coefficients.
As a result, solely the classically known self-heating terms are contained
in the model and all additional, transient heating mechanisms are
excluded:
\begin{equation}
H  =\frac{1}{\sigma_{n}}\left\Vert \mathbf{j}_{n}\right\Vert ^{2}+\frac{1}{\sigma_{p}}\left\Vert \mathbf{j}_{p}\right\Vert ^{2}-T\,\mathbf{j}_{n}\cdot\nabla P_{n}-T\,\mathbf{j}_{p}\cdot\nabla P_{p}+q\left(\varphi_{p}-\varphi_{n}+\Pi_{p}-\Pi_{n}\right)R.\label{eq: heat source final}
\end{equation}

Finally, we rewrite the recombination heating term using the Seebeck
coefficients (\ref{eq: explicit Seebeck-1}) and Eq.~(\ref{eq: carrier density state equations}).
One obtains
\begin{equation*}
H_{R}=\left(E_{g}\left(T\right)-TE_{g}^{\prime}\left(T\right)+\left[\frac{TN_{v}^{\prime}\left(T\right)}{N_{v}\left(T\right)}g\left(\frac{p}{N_{v}\left(T\right)}\right)+\frac{TN_{c}^{\prime}\left(T\right)}{N_{c}\left(T\right)}g\left(\frac{n}{N_{c}\left(T\right)}\right)\right]k_{B}T\right)R.
\end{equation*}
The last term  describes the (differential) average thermal
energy per recombining electron-hole pair. For an effective density
of states function $N_{c/v}\propto T^{3/2}$ and non-degenerate carrier
statistics, we recover the classical result $H_{R}\approx\left(E_{g}\left(T\right)-TE_{g}^{\prime}\left(T\right)+3k_{B}T\right)R.$
This yields a clear interpretation of the degeneracy factor $g$ (see
Eq.~(\ref{eq: degeneracy factor - both expressions})): It describes
the increased average thermal energy of the Fermi gas due to Pauli
blocking in comparison to the non-degenerate case at the same carrier
density. We emphasize that the Kelvin formula immediately yields the
correct average kinetic energy $2\times\frac{3}{2}k_{B}T$ of the
three-dimensional electron-hole plasma just from the temperature dependency
of the effective density of states function $N_{c/v}\propto T^{3/2}$.
This does in general not hold for Seebeck coefficients derived from
the Boltzmann transport equation in relaxation time approximation,
where the average thermal energy of the electron-hole plasma in the
recombination heat term depends on a scattering parameter, see e.\,g.
Ref.~\citep{Wenzel2017}.

The dissipated heat is closely related with the electrical power injected through the contacts.
The  global power balance equation for the present model is derived in \ref{sec: power balance}.

\subsection{Electrical current densities in drift-diffusion form\label{sec: Drift-diffusion-current-densities}}

In this section we recast the electrical current density expressions
from the thermodynamic form (\ref{eq: current densities-2}) to the
drift-diffusion form. As we will see below, the Kelvin formula for
the Seebeck coefficient allows to entirely absorb the thermally driven
part of the electrical current density in the diffusion coefficient
via a generalized Einstein relation. Thus, the $\nabla T$ term can
be eliminated in the drift-diffusion form, which significantly simplifies
the current density expression. Our derivation is based on rewriting
the gradient of the quasi-Fermi potential using the free energy density
(\ref{eq: free energy density-1}) and further thermodynamic relations
stated above. In the following, we sketch the essential steps for
the electron current density, the corresponding expression for the
holes follows analogously. We obtain
\[
-q\nabla\varphi_{n}\stackrel{\text{Eq.}\,\eqref{eq: conjugate fields-1}}{=}\nabla\frac{\partial f}{\partial n}\stackrel{\text{Eq.}\,\eqref{eq: free energy density-1}}{=}\nabla\left(\frac{\partial f_{\text{Coul}}}{\partial n}+\frac{\partial f_{\text{e--h}}}{\partial n}\right)\stackrel{\text{Eq.}\,\eqref{eq: electrostatic derivatives-1}}{=}-q\nabla\phi+\frac{\partial^{2}f_{\text{e--h}}}{\partial n^{2}}\nabla n+\frac{\partial^{2}f_{\text{e--h}}}{\partial n\,\partial p}\nabla p+\frac{\partial^{2}f_{\text{e--h}}}{\partial n\,\partial T}\nabla T,
\]
where we have separated the contributions from the Coulomb interaction
energy $f_{\text{Coul}}$ (leading to drift in the electric field)
and the quasi-free electron-hole plasma (yielding Hessian matrix elements
of the ideal Fermi gas' free energy density $f_{\text{e--h}}$). The
electrons and holes are decoupled in the non-interacting Fermi-gas
such that
\begin{align*}
\frac{\partial^{2}f_{\text{e--h}}}{\partial n\,\partial p} & =0.
\end{align*}
Moreover, since (i)~the Coulomb interaction energy is independent
of the temperature and therefore does not contribute to the system's
entropy and (ii)~the lattice contribution $f_{L}$ is independent
of the carrier densities, it holds
\[
\frac{\partial^{2}f_{\text{e--h}}}{\partial n\,\partial T}=-\frac{\partial s}{\partial n},
\]
where $s$ is the entropy density of the full system (see the last
formula in Eq.~(\ref{eq: conjugate fields-1})). Thus, we arrive
at
\[
\nabla\varphi_{n}=\nabla\phi-\frac{1}{q}\frac{\partial^{2}f_{\text{e--h}}}{\partial n^{2}}\nabla n+\frac{1}{q}\frac{\partial s}{\partial n}\nabla T,
\]
which must be substituted in Eq.~(\ref{eq: current densities-2})
to obtain
\[
\mathbf{j}_{n}=-\sigma_{n}\left(\nabla\phi-\frac{1}{q}\frac{\partial^{2}f_{\text{e--h}}}{\partial n^{2}}\nabla n+\left[\frac{1}{q}\frac{\partial s}{\partial n}+P_{n}\right]\nabla T\right)=-\sigma_{n}\nabla\phi+\sigma_{n}\frac{1}{q}\frac{\partial^{2}f_{\text{e--h}}}{\partial n^{2}}\nabla n.
\]
In the last step, we have used the Kelvin formula (\ref{eq: Kelvin formula})
for the Seebeck coefficient. The temperature gradient term vanishes
exactly, since reversing the order of the derivatives in the Hessian
of the free energy density immediately yields the definition (\ref{eq: Kelvin formula})
and cancels with the Seebeck term in Eq.~\eqref{eq: current densities-2}.
The same result can be obtained by simply inverting the carrier density
state equation (\ref{eq: carrier density state equations}) and using
the explicit expression (\ref{eq: explicit Seebeck-1}). With the
electrical conductivities $\sigma_{n}=qM_{n}n$, $\sigma_{p}=qM_{p}p$
and
\begin{align*}
\frac{\partial^{2}f_{\text{e--h}}}{\partial n^{2}} & =\frac{k_{B}T}{n}g\left(\frac{n}{N_{c}\left(T\right)}\right), & \frac{\partial^{2}f_{\text{e--h}}}{\partial p^{2}} & =\frac{k_{B}T}{p}g\left(\frac{p}{N_{v}\left(T\right)}\right),
\end{align*}
(from Eq.~(\ref{eq: non-interacting Fermi gas chemical potentials-1})),
we finally arrive at the drift-diffusion form:
\begin{align}
\mathbf{j}_{n} & =-qM_{n}n\nabla\phi+qD_{n}\left(n,T\right)\nabla n, & \mathbf{j}_{p} & =-qM_{p}p\nabla\phi-qD_{p}\left(p,T\right)\nabla p.\label{eq: drift-diffusion currents}
\end{align}
The diffusion coefficients are given by the \emph{generalized} Einstein
relation \citep{Landsberg1952,VanVliet1976}
\begin{align}
D_{n}\left(n,T\right) & =\frac{k_{B}TM_{n}}{q}g\left(\frac{n}{N_{c}\left(T\right)}\right), & D_{p}\left(p,T\right) & =\frac{k_{B}TM_{p}}{q}g\left(\frac{p}{N_{v}\left(T\right)}\right).\label{eq: generalized Einstein relations-1}
\end{align}
Here the degeneracy factor $g$ describes an effective enhancement
of the diffusion current that depends nonlinearly on the carrier densities,
which results from the increased average thermal energy of the carriers
in the case of Fermi--Dirac statistics (see above). The diffusion
enhancement due to carrier degeneracy has been found to be important
in, e.\,g., semiconductor laser diodes \citep{Shore1976}, quantum-photonic devices operated at cryogenic temperatures \cite{Kantner2016a,Kantner2016}
as well as organic field-effect transistors \citep{Roichman2002} and light emitting
diodes \citep{Mensfoort2008}. We emphasize that the drift-diffusion
form (\ref{eq: drift-diffusion currents}) of the current densities
is fully equivalent to the thermodynamic form (\ref{eq: current densities-2}).
Thus, even though the $\nabla T$ term is eliminated, the thermoelectric
cross-coupling via the Seebeck effect is fully taken into account
via the temperature dependency of the diffusion coefficient.
A generalization to the case of hot carrier transport (with multiple temperatures) is described in \ref{sec: Generalization to the case of multiple temperatures}.
For Seebeck coefficients that deviate from the Kelvin formula, additional
thermodiffusion terms emerge.

\section{Non-isothermal generalization of the Scharfetter--Gummel scheme
for degenerate semiconductors \label{sec: Non-isothermal generalization of the Scharfetter=002013Gummel scheme for degenerate semiconductors}}

The typically exponentially varying carrier densities in semiconductor
devices lead to numerical instabilities when using a standard finite
difference discretization. In particular, the naive discretization
approach results in spurious oscillations and may cause unphysical
results such as negative carrier densities \citep{Brezzi1989,Farrell2017}.
A robust discretization scheme for the drift-diffusion current density
was introduced by Scharfetter and Gummel \citep{Scharfetter1969},
who explicitly solved the current density expressions as a separate
differential equation along the edge between two adjacent nodes of
the mesh. The resulting discretized current density expressions feature
exponential terms that reflect the characteristics of the doping profile
and allow for numerically stable calculations. Over the last decades,
several generalizations of the Scharfetter--Gummel method have been
proposed for either degenerate semiconductors \citep{Yu1985,Gajewski1993,Bessemoulin-Chatard2012,Koprucki2013,Koprucki2015,Fuhrmann2015}
or non-isothermal carrier transport with included thermoelectric cross
effects \citep{Tang1984,McAndrew1985,Rudan1986,Forghieri1988,Chen1991,Souissi1991,TenThijeBoonkkamp1993,Smith1993}.

In this section, we derive two different generalizations of the Scharfetter--Gummel
scheme for degenerate semiconductors obeying the Kelvin formula for
the Seebeck coefficient. Both schemes differ in the treatment of degeneration
effects and are obtained by extending the approaches previously developed
in Refs.~\citep{Bessemoulin-Chatard2012,Koprucki2015} and \citep{Yu1985}.
First, we outline the finite volume method in Sec.~\ref{sec: finite volume discretization}
and then introduce the non-isothermal Scharfetter--Gummel schemes
in Sec.~\ref{sec: Generalized Scharfetter=002013Gummel scheme}.
We study important limiting cases and structure preserving properties
of the discretizations (Sec.~\ref{Sec: Limiting cases and structure preserving properties}),
give a detailed comparison with the numerically exact solution of
the underlying two-point boundary value problem (Sec.~\ref{sec: Comparison with numerically exact solution})
and derive analytical error bounds (Sec.~\ref{sec: Analytical error estimate}).
Finally, we present a numerical convergence analysis by means of numerical
simulations of a one-dimensional p-n-diode in Sec.~\ref{sec: benchmark simulation}.

\subsection{Finite volume discretization \label{sec: finite volume discretization}}

\begin{figure}
\centering

\includegraphics{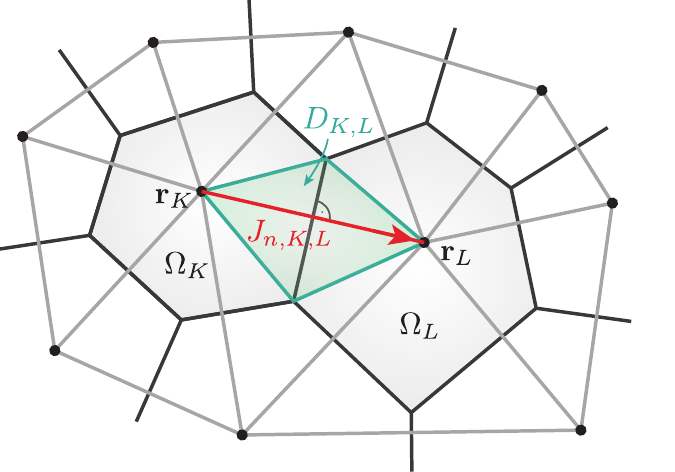}

\caption{Delaunay triangulation and construction of Vorono\"i cells. The red
arrow indicates the discrete current $J_{n,K,L}$ between two neighboring
control volumes $\Omega_{K}$ and $\Omega_{L}$. The green area is
the bi-hyperpyramid (or ``diamond cell'') $D_{K,L}$ with height $\left\Vert \mathbf{r}_{L}-\mathbf{r}_{K}\right\Vert $
and internal face $\vert\partial\Omega_{L}\cap\partial\Omega_{K}\vert$.
Adapted, with permission, from Ref.~\cite{Kantner2019a}. \copyright~2019 IEEE.
}

\label{fig: Voronoii}
\end{figure}

We assume a boundary conforming Delaunay triangulation \citep{Si2010}
of the point set $\mathbf{R}=\left\{ \mathbf{r}_{K}\right\} _{K=1\ldots N_{\text{nodes}}}$,
$\mathbf{r}_{K}\in\Omega$, where $\Omega\subset\mathbb{R}^{d}$ is
the computational domain with dimensionality $d=\{ 1,2,3 \}$. The dual mesh is
given by the Vorono\"i cells
\[
\Omega_{K}=\left\{ \mathbf{r}\in\Omega:\left\Vert \mathbf{r}-\mathbf{r}_{K}\right\Vert \leq\left\Vert \mathbf{r}-\mathbf{r}_{L}\right\Vert \text{ for all }\mathbf{r}_{L}\in\mathbf{R}\text{ with }\mathbf{r}_{L}\neq \mathbf{r}_{K}\right\} ,
\]
which provides a non-overlapping tessellation $\Omega=\bigcup_{K}\Omega_{K}$
of the domain. This represents an admissible mesh in the sense of
Ref.~\citep{Eymard2000}. The finite volume discretization of the
system (\ref{eq: Poisson equation})--(\ref{eq: heat equation})
is obtained by integration over the cell $\Omega_{K}$ and usage
of the divergence theorem \citep{Eymard2000,Farrell2017}. The discrete
(stationary) non-isothermal drift-diffusion system reads\begin{subequations}\label{eq: discrete energy-drift-diffusion system}
\begin{align}
-\sum_{L\in\mathcal{N}\left(K\right)}s_{K,L}\varepsilon\left(\phi_{L}-\phi_{K}\right) & =q\vert\Omega_{K}\vert\left(C_{K}+p_{K}-n_{K}\right)\label{eq: discrete Poisson equation}\\
-\sum_{L\in\mathcal{N}\left(K\right)}s_{K,L}J_{n,K,L} & =-q\vert\Omega_{K}\vert R_{K},\label{eq: discrete electron transport equation}\\
+\sum_{L\in\mathcal{N}\left(K\right)}s_{K,L}J_{p,K,L} & =-q\vert\Omega_{K}\vert R_{K},\label{eq: discrete hole transport equation}\\
-\sum_{L\in\mathcal{N}\left(K\right)}s_{K,L}\kappa\left(T_{L}-T_{K}\right) & =\frac{1}{2}\sum_{L\in\mathcal{N}\left(K\right)}s_{K,L}\left(H_{J,K,L}+H_{\text{T--P},K,L}\right)+\vert\Omega_{K}\vert H_{R,K}\label{eq: discrete heat equation}
\end{align}
\end{subequations}
with the flux projections
\begin{align}
J_{n,K,L} & =\left(\mathbf{r}_{L}-\mathbf{r}_{K}\right)\cdot\mathbf{j}_{n}, & J_{p,K,L} & =\left(\mathbf{r}_{L}-\mathbf{r}_{K}\right)\cdot\mathbf{j}_{p} \label{eq: discrete current normal projection}
\end{align}
on the edge $\overline{KL} := \{ x \, \mathbf{r}_L + (1-x)\, \mathbf{r}_K \, \vert\,  x\in\mathbb{R}, 0\leq x \leq 1 \}$.
The geometric factors in Eq.~(\ref{eq: discrete energy-drift-diffusion system})
are the volume $\vert\Omega_{K}\vert$ of the $K$-th Vorono\"i cell
and the edge factor 
\begin{equation}
s_{K,L}=\frac{\vert\partial\Omega_{K}\cap\partial\Omega_{L}\vert}{\left\Vert \mathbf{r}_{L}-\mathbf{r}_{K}\right\Vert }. \label{eq: edge-factor}
\end{equation}
The symbol $\mathcal{N}\left(K\right)$ denotes the set of nodes adjacent
to $K$. For the sake of simplicity we restrict ourselves to the case
of a homogeneous material. This limitation is not important for
the flux discretization, as the discrete fluxes appear only along
possible heterointerfaces (edges of the primary simplex grid, see
Fig.~(\ref{fig: Voronoii})) but never across. In the case of heterostructures,
the currents along material interfaces are weighted by the respective
edge factors.
Moreover, boundary terms on $\partial\Omega\cap\Omega_{K}\neq\emptyset$
are omitted in Eq.~(\ref{eq: discrete energy-drift-diffusion system}), which are treated in the standard way as described in Ref.~\cite{Eymard2000} and references therein.
The discrete electron density reads $n_{K}=N_{c}\left(T_{K}\right)\mathscr{F}\left(\eta_{n,K}\right)$
with $\eta_{n,K}=-\left(E_{c}\left(T_{K}\right)-q\phi_{K}+q\varphi_{n,K}\right)/\left(k_{B}T_{K}\right)$
(holes analogously).
The discrete recombination rate $R_K$ is obtained by locally evaluating Eq.~\eqref{eq: recombination rate} as
$R_{K}=R(\phi_{K},\varphi_{n,K},\varphi_{p,K},T_{K})$. Similarly, the discrete doping density
on $\Omega_K$ is taken as $C_K = C(\mathbf{r}_K)$. The discrete self-heating terms are
\begin{subequations}\label{eq: discrete heat generation rate}
\begin{align}
H_{J,K,L} & =-J_{n,K,L}\left(\varphi_{n,L}-\varphi_{n,K}+P_{n,K,L}\left(T_{L}-T_{K}\right)\right)-J_{p,K,L}\left(\varphi_{p,L}-\varphi_{p,K}+P_{p,K,L}\left(T_{L}-T_{K}\right)\right),\label{eq: discrete Joule heating}\\
H_{\text{T--P},K,L} & =-T_{K,L}J_{n,K,L}\left(P_{n,L}-P_{n,K}\right)-T_{K,L}J_{p,K,L}\left(P_{p,L}-P_{p,K}\right),\label{eq: discrete Thomson Peltier heating}\\
H_{R,K} & =q\left(\varphi_{p,K}-\varphi_{n,K}+T_{K}\left(P_{p,K}-P_{n,K}\right)\right)R_{K}.\label{eq: discrete recombination heating}
\end{align}
\end{subequations}
The finite volume discretization of the Joule and Thomson--Peltier heating terms is not straightforward \citep{Bradji2008,Chainais-Hillairet2009, Eymard2003}. Details on the derivation of Eqs.~\eqref{eq: discrete Joule heating}--\eqref{eq: discrete Thomson Peltier heating} are provided in \ref{sec: Discretization of the heat source term}.
The discretization of the edge current densities $J_{n/p,K,L}$,
the edge-averaged temperature $T_{K,L}$ and the Seebeck coefficients
$P_{n/p,K,L}$ along the edge $\overline{KL}$ are subject to the
following sections.

\subsection{Discretization of the current density expression\label{sec: Generalized Scharfetter=002013Gummel scheme}}

The discretization of $J_{n/p,K,L}$ is obtained
by integrating the current density expressions (\ref{eq: drift-diffusion currents})
along the edge $\overline{KL}$ between two adjacent nodes of the
mesh. Since the Kelvin formula implies a remarkably simple form of
the electrical current densities in drift-diffusion form, where the
thermal driving force is eliminated exactly (see Sec.~\ref{sec: Drift-diffusion-current-densities}),
this allows for a straightforward adaptation of the Scharfetter--Gummel
schemes developed for the isothermal case. We assume the electrostatic
field $\mathbf{E}=-\nabla\phi$ and the temperature gradient $\nabla T$
to be constant along the edge $\overline{KL}$, such that
\begin{align*}
\phi\left(x\right) & =x\,\phi_{L}+\left(1-x\right)\phi_{K}, & T\left(x\right) & =x\,T_{L}+\left(1-x\right)T_{K},
\end{align*}
where $x\in\left[0,1\right]$ parametrizes the coordinate on the edge
$\mathbf{r}\left(x\right)=x\thinspace\mathbf{r}_{L}+\left(1-x\right)\mathbf{r}_{K}$.
Tacitly, these assumptions have already been used above in Eqs.~(\ref{eq: discrete Poisson equation})
and (\ref{eq: discrete heat equation}). Moreover, also the mobilities
$M_{n/p}$ and the fluxes $J_{n/p,K,L}$ are assumed to be constant
on the edge. For the electron current density, this yields the two-point
boundary value problem (BVP)
\begin{align}
k_{B}T\left(x\right)g\left(\frac{n\left(x\right)}{N_{c}\left(T\left(x\right)\right)}\right)\frac{\mathrm{d}n}{\mathrm{d}x} & =q\left(\phi_{L}-\phi_{K}\right)n\left(x\right)+\frac{J_{n,K,L}}{M_{n}}, & n\left(0\right) & =n_{K}, & n\left(1\right) & =n_{L},\label{eq: two-point boundary value problem}
\end{align}
on $x=\left[0,1\right]$. The problem for the holes current density
is analogous.

In the non-degenerate case (Maxwell--Boltzmann statistics) the degeneracy
factor is exactly $g\equiv1$, such that the problem can be solved
exactly by separation of variables. One obtains
\begin{align*}
\int_{n_{K}}^{n_{L}}\frac{\mathrm{d}n}{\frac{J_{n,K,L}}{qM_{n}\left(\phi_{L}-\phi_{K}\right)}+n} & =\frac{q\left(\phi_{L}-\phi_{K}\right)}{k_{B}}\int_{0}^{1}\frac{\mathrm{d}x}{T\left(x\right)},
\end{align*}
where the integral on the right hand side yields the (inverse) logarithmic
mean temperature
\begin{align}
\int_{0}^{1}\frac{\mathrm{d}x}{T\left(x\right)} & =\int_{0}^{1}\frac{\mathrm{d}x}{x\,T_{L}+\left(1-x\right)T_{K}}=\frac{1}{\Lambda\left(T_{L},T_{K}\right)}\equiv\frac{1}{T_{K,L}}, & \Lambda\left(x,y\right) & =\frac{x-y}{\log{\left(x/y\right)}},\label{eq: logarithmic mean temperature}
\end{align}
where $\Lambda\left(x,y\right)$ is the logarithmic mean. Solving for
the flux yields the non-isothermal Scharfetter--Gummel scheme
\begin{align}
J_{n,K,L}^{\text{ndeg}} & =M_{n}k_{B}T_{K,L}\left(n_{L}B\left(X_{n,K,L}^{\text{ndeg}}\right)-n_{K}B\left(-X_{n,K,L}^{\text{ndeg}}\right)\right),\qquad X_{n,K,L}^{\text{ndeg}}=\frac{q\left(\phi_{L}-\phi_{K}\right)}{k_{B}T_{K,L}},\label{eq: non-degenerate, non-isothermal SG}
\end{align}
where $B\left(x\right)=x/\left(\exp{\left(x\right)}-1\right)$ is
the Bernoulli function. The Bernoulli function is closely related
to the logarithmic mean: $B\left(x\right)=1/\Lambda\left(\exp{(x)},1\right)$.
At isothermal conditions $T_{K} \equiv T_{L}$, Eq.~(\ref{eq: non-degenerate, non-isothermal SG})
reduces to the original Scharfetter--Gummel scheme \citep{Scharfetter1969}.

In the case of Fermi--Dirac statistics ($g\neq1$), no closed-form
solution exists such that approximate solutions of the BVP (\ref{eq: two-point boundary value problem})
are required. As the degeneracy factor $g$ depends on both the carrier
density and temperature, the problem is not even separable

\subsubsection{Modified thermal voltage scheme \label{sec: Modified thermal voltage scheme}}

Following Refs.~\citep{Bessemoulin-Chatard2012,Koprucki2015}, we
solve the BVP (\ref{eq: two-point boundary value problem}) by freezing
the degeneracy factor $g\left(n/N_{c}\left(T\right)\right)\to g_{n,K,L}$
to a carefully chosen average. The resulting problem has the same
structure as in the non-degenerate case (see above), but with a modified
thermal voltage $k_{B}T_{K,L}/q\to k_{B}T_{K,L}g_{n,K,L}/q$ along
the edge, which takes the temperature variation and the degeneration
of the electron gas into account. This yields the \emph{modified thermal
voltage scheme}\begin{subequations}\label{eq: log mean temp Chatard scheme}
\begin{align}
J_{n,K,L}^{g} & =M_{n}k_{B}T_{K,L}g_{n,K,L}\left(n_{L}B\left(X_{n,K,L}^{g}\right)-n_{K}B\left(-X_{n,K,L}^{g}\right)\right),\qquad X_{n,K,L}^{g}=\frac{q\left(\phi_{L}-\phi_{K}\right)}{k_{B}T_{K,L}g_{n,K,L}},\label{eq: generalized Chatard}
\end{align}
where $T_{K,L}$ is the logarithmic mean temperature (\ref{eq: logarithmic mean temperature}).
In order to ensure the consistency with the thermodynamic equilibrium
and boundedness $g_{n,K}\leq g_{n,K,L}\leq g_{n,L}$ (for $\eta_{n,K}\leq\eta_{n,L}$
or with $K\leftrightarrow L$ else), the edge-averaged degeneracy
factor is taken as \citep{Bessemoulin-Chatard2012,Koprucki2015}
\begin{equation}
g_{n,K,L}=\frac{\eta_{n,L}-\eta_{n,K}}{\log{\left(\mathscr{F}\left(\eta_{n,L}\right)/\mathscr{F}\left(\eta_{n,K}\right)\right)}}=\frac{\mathscr{F}^{-1}\left(n_{L}/N_{c}\left(T_{L}\right)\right)-\mathscr{F}^{-1}\left(n_{K}/N_{c}\left(T_{K}\right)\right)}{\log{\left(n_{L}/N_{c}\left(T_{L}\right)\right)}-\log{\left(n_{K}/N_{c}\left(T_{K}\right)\right)}}.\label{eq: average degeneracy factor}
\end{equation}
\end{subequations}In the limit of $\eta_{n,L}=\eta_{n,K}$, it approaches
the common nodal value 
\[
\lim_{\eta_{n,L}\to\eta_{n,K}\equiv\bar{\eta}_{n}}g_{n,K,L}=\frac{\mathscr{F}\left(\bar{\eta}_{n}\right)}{\mathscr{F}^{\prime}\left(\bar{\eta}_{n}\right)}=g\left(\mathscr{F}\left(\bar{\eta}_{n}\right)\right)\equiv g_{\eta}\left(\bar{\eta}_{n}\right).
\]
For constant temperature $T_{L}=T_{K}=T$, the scheme reduces to the
modified Scharfetter--Gummel scheme discussed in Refs.~\citep{Bessemoulin-Chatard2012,Koprucki2015}.
It can thus be regarded as a non-isothermal generalization of this
approach. In the non-degenerate limit $g=1$, it reduces to the non-isothermal,
non-degenerate Scharfetter--Gummel scheme (\ref{eq: non-degenerate, non-isothermal SG}).

\subsubsection{Modified drift scheme \label{sec: Correction factor scheme}}

The traditional approach for the inclusion of degeneration effects
in the Scharfetter--Gummel scheme, that is widely used in commercial
software packages, is based on introducing the correction factors
\citep{Yu1985,Synopsys2010,Silvaco2016}
\begin{equation}
\gamma\left(\eta\right)=\frac{\mathscr{F}\left(\eta\right)}{\exp\left(\eta\right)},\label{eq: correction factor-1-1}
\end{equation}
and rearranging the current density expression with nonlinear diffusion
(\ref{eq: drift-diffusion currents}) (involving the generalized Einstein
relation (\ref{eq: generalized Einstein relations-1})) into a form
with linear diffusion and a modified drift term:
\begin{align}
\mathbf{j}_{n} & =\sigma_{n}\mathbf{E}_{n}+M_{n}k_{B}T\nabla n+\frac{k_{B}}{q}\sigma_{n}\thinspace\rho_{n}\left(T,\eta_{n}\right)\nabla T, & \mathbf{E}_{n} & =-\nabla\left(\phi+\frac{k_{B}T}{q}\log\left(\gamma\left(\eta_{n}\right)\right)\right).\label{eq: drift-diffusion form with linear diffusion}
\end{align}
Here, the degeneration of the electron gas induces a thermodiffusion
term with the coefficient
\begin{equation}
\rho_{n}\left(T,\eta_{n}\right)=\log\left(\gamma\left(\eta_{n}\right)\right)-\frac{TN_{c}^{\prime}\left(T\right)}{N_{c}\left(T\right)}\frac{\gamma^{\prime}\left(\eta_{n}\right)/\gamma\left(\eta_{n}\right)}{1+\gamma^{\prime}\left(\eta_{n}\right)/\gamma\left(\eta_{n}\right)}=\log\left(\gamma\left(\eta_{n}\right)\right)+\frac{TN_{c}^{\prime}\left(T\right)}{N_{c}\left(T\right)}\left(g_{\eta}\left(\eta_{n}\right)-1\right),\label{eq: deviation function}
\end{equation}
that vanishes exactly in the non-degenerate limit ${\gamma\left(\eta\right)\equiv1\equiv g_{\eta}\left(\eta\right)}$.
Hence, the function $\rho_{n}\left(T,\eta_{n}\right)$ quantifies
the difference between the degenerate and the non-degenerate Seebeck-coefficient,
see Fig.~\ref{fig: Seebeck coefficient-2}\,(a). On the step from
Eq.~(\ref{eq: drift-diffusion currents}) to (\ref{eq: drift-diffusion form with linear diffusion}),
we have used the relation $g_{\eta}\left(\eta\right)=\left(1+\gamma^{\prime}\left(\eta\right)/\gamma\left(\eta\right)\right)^{-1}$.
A plot of the correction factor (\ref{eq: correction factor-1-1})
is given in Fig.~\ref{fig: distribution function and degeneracy factor}\,(c).

The current density expression (\ref{eq: drift-diffusion form with linear diffusion})
is discretized by projecting the current on the edge $\overline{KL}$,
assuming the effective electric field $\mathbf{E}_{n}$ to be a constant
along the edge, and freezing $\rho_{n}\left(T,\eta_{n}\right)\to\rho_{n,K,L}$
to a constant average value.
Here, different averages can be taken for $\rho_{n,K,L}$, see Fig.\,\ref{fig: SG comparison 1D plots}\,(c). The influence of this choice will be discussed below in Sec.~\ref{sec: Comparison with numerically exact solution}.
Along the same lines as above, one arrives
at the \emph{modified drift scheme}\begin{subequations}\label{eq: modified drift scheme}
\begin{align}
\begin{aligned}J_{n,K,L}^{\gamma}\end{aligned}
 & =M_{n}k_{B}T_{K,L}\left(n_{L}B\left(X_{n,K,L}^{\gamma}\right)-n_{K}B\left(-X_{n,K,L}^{\gamma}\right)\right),\label{eq: modified advection scheme-1}
\end{align}
with 
\begin{equation}
X_{n,K,L}^{\gamma}=\frac{q\left(\phi_{L}-\phi_{K}\right)}{k_{B}T_{K,L}}+\frac{T_{L}\log\left(\gamma\left(\eta_{n,L}\right)\right)-T_{K}\log\left(\gamma\left(\eta_{n,K}\right)\right)}{T_{K,L}}-\rho_{n,K,L}\log{\left(\frac{T_{L}}{T_{K}}\right)}.\label{eq: modified advection scheme-1 X}
\end{equation}
\end{subequations}Again, $T_{K,L}$ is the logarithmic mean temperature
(\ref{eq: logarithmic mean temperature}). The corresponding non-degenerate
limit \eqref{eq: non-degenerate, non-isothermal SG} is easily recovered by $\gamma\left(\eta_{n,L/K}\right)\to1$ and $\rho_{n,K,L}\to0$.

\subsection{Limiting cases and structure preserving properties\label{Sec: Limiting cases and structure preserving properties}}

In the following, we investigate some important limiting cases and
structure preserving properties of the generalized Scharfetter--Gummel
schemes (\ref{eq: log mean temp Chatard scheme}) and (\ref{eq: modified drift scheme}).
This includes an analysis of the consistency of the discrete expressions
with fundamental thermodynamical principles (thermodynamic equilibrium,
second law of thermodynamics). To this end, it is convenient to rewrite
both expressions using the identity $B\left(-x\right)=\exp{\left(x\right)}\,B\left(x\right)$
and the logarithmic mean $\Lambda$ (see Eq.~(\ref{eq: logarithmic mean temperature}))
as
\begin{align}
J_{n,K,L}^{g} & =-\sigma_{n,K,L}^{g}\left(\varphi_{n,L}-\varphi_{n,K}+P_{n,K,L}^{g}\left(T_{L}-T_{K}\right)\right) & \text{with}\quad\sigma_{n,K,L}^{g}= & qM_{n}\frac{\Lambda\left(n_{L}\exp{\left(-\frac{1}{2}X_{n,K,L}^{g}\right)},n_{K}\exp{\left(\frac{1}{2}X_{n,K,L}^{g}\right)}\right)}{\mathrm{sinhc}{\left(\frac{1}{2}X_{n,K,L}^{g}\right)}},\label{eq: modified thermal voltage scheme - discrete thermodynamic form}
\end{align}
and
\begin{align}
J_{n,K,L}^{\gamma} & =-\sigma_{n,K,L}^{\gamma}\left(\varphi_{n,L}-\varphi_{n,K}+P_{n,K,L}^{\gamma}\left(T_{L}-T_{K}\right)\right) & \text{with}\quad\sigma_{n,K,L}^{\gamma}= & qM_{n}\frac{\Lambda\left(n_{L}\exp{\left(-\frac{1}{2}X_{n,K,L}^{\gamma}\right)},n_{K}\exp{\left(\frac{1}{2}X_{n,K,L}^{\gamma}\right)}\right)}{\mathrm{sinhc}{\left(\frac{1}{2}X_{n,K,L}^{\gamma}\right)}},\label{eq: modified drift scheme - discrete thermodynamic form}
\end{align}
where $\mathrm{sinhc}\left(x\right)=\sinh{\left(x\right)}/x$. This
representation directly corresponds to the continuous current density
expression in the thermodynamic form (\ref{eq: current densities-2}),
where the conductivity along the edge $\sigma_{n,K,L}^{g/\gamma}$
is determined by a ``tilted'' logarithmic average of the nodal carrier
densities. Both expressions (\ref{eq: modified thermal voltage scheme - discrete thermodynamic form})--(\ref{eq: modified drift scheme - discrete thermodynamic form})
have a common structure, but differ in the discrete conductivity $\sigma_{n,K,L}^{g}\neq\sigma_{n,K,L}^{\gamma}$
(due to $X_{n,K,L}^{g}\neq X_{n,K,L}^{\gamma}$) and the discrete
Seebeck coefficients $P_{n,K,L}^{g}\neq P_{n,K,L}^{\gamma}$ along
the edge, which are implicitly prescribed by the Scharfetter--Gummel discretization
procedure. The latter read
\begin{subequations}\label{eq: edge averaged Seebeck coefficients}
\begin{align}
P_{n,K,L}^{g} & =-\frac{k_{B}}{q}\left[\log{\left(\frac{N_{c}\left(T_{L}\right)}{N_{c}\left(T_{K}\right)}\right)}\frac{g_{n,K,L}}{\log{\left(T_{L}/T_{K}\right)}}-\frac{\left(T_{L}-T_{K,L}\right)\eta_{n,L}-\left(T_{K}-T_{K,L}\right)\eta_{n,K}}{T_{L}-T_{K}}-\frac{1}{k_{B}}\frac{E_{c}\left(T_{L}\right)-E_{c}\left(T_{K}\right)}{T_{L}-T_{K}}\right]\label{eq: edge averaged Seebeck coefficients - mod thermal voltage}
\end{align}
and
\begin{align}
P_{n,K,L}^{\gamma} & =-\frac{k_{B}}{q}\Bigg[\log{\left(\frac{N_{c}\left(T_{L}\right)}{N_{c}\left(T_{K}\right)}\right)}\frac{1}{\log{\left(T_{L}/T_{K}\right)}}+\rho_{n,K,L}-\frac{T_{L}\log\left(\gamma\left(\eta_{n,L}\right)\right)-T_{K}\log\left(\gamma\left(\eta_{n,K}\right)\right)-T_{K,L}\log{\left(\gamma\left(\eta_{n,L}\right)/\gamma\left(\eta_{n,K}\right)\right)}}{T_{L}-T_{K}}\label{eq: edge averaged Seebeck coefficients - correction factor}\\
 & \phantom{=-\frac{k_{B}}{q}\Bigg[\log{\left(\frac{N_{c}\left(T_{L}\right)}{N_{c}\left(T_{K}\right)}\right)}\frac{1}{\log{\left(T_{L}/T_{K}\right)}}+\rho_{n,K,L}}-\frac{\left(T_{L}-T_{K,L}\right)\eta_{n,L}-\left(T_{K}-T_{K,L}\right)\eta_{n,K}}{T_{L}-T_{K}}-\frac{1}{k_{B}}\frac{E_{c}\left(T_{L}\right)-E_{c}\left(T_{K}\right)}{T_{L}-T_{K}}\Bigg].\nonumber 
\end{align}
\end{subequations}
The discrete Seebeck coefficients (\ref{eq: edge averaged Seebeck coefficients})
enter the discrete Joule heat term (\ref{eq: discrete Joule heating})
and thus the discrete entropy production rate (see Sec.~\ref{sec: Non-negativity-of-the discrete dissipation rate}
below). Out of the thermodynamic equilibrium, the discrete Seebeck coefficients
(\ref{eq: edge averaged Seebeck coefficients}) determine the point
of compensating (discrete) chemical and thermal current flow such that $J_{n,K,L}\vert_{\varphi_{n,K}\neq\varphi_{n,L},T_{K}\neq T_{L}}=0$.
In other words, there is a non-equilibrium configuration with $T_K \neq T_L$ and $\varphi_{n,K}\neq\varphi_{n,L}$, 
where the discrete Seebeck coefficient $P_{n,K,L}^{g/\gamma} = - \left(\varphi_{n,L} - \varphi_{n,K}\right)/\left(T_L - T_K\right)$
equals the (negative) ratio of both discrete driving forces such that the discrete current density is zero.
This compensation point is in general slightly different between both
schemes (see inset of Fig.~\ref{fig: SG comparison 1D plots}\,(c)).
In the limit of a small temperature gradient and a small difference
in the reduced Fermi energy along the edge, both discrete Seebeck
coefficients approach in leading order the continuous expression (\ref{eq: explicit Seebeck - electrons-1})
\begin{equation*}
P_{n,K,L}^{g/\gamma} =-\frac{k_{B}}{q}\left[\frac{\bar{T}N_{c}^{\prime}(\bar{T})}{N_{c}(\bar{T})}g_{\eta}\left(\bar{\eta}_{n}\right)-\bar{\eta}_{n}-\frac{1}{k_{B}}E_{c}^{\prime}(\bar{T})\right]+\mathcal{O}(\delta\eta_{n}^{2})+\mathcal{O}(\delta\eta_{n}\,\delta\Theta)+\mathcal{O}(\delta\Theta^{2}),
\end{equation*}
where $\delta\Theta=\left(T_{L}-T_{K}\right)/\bar{T}$, $\bar{T}=\frac{1}{2}\left(T_{L}+T_{K}\right)$,
$\delta\eta_{n}=\eta_{n,L}-\eta_{n,K}$ and $\bar{\eta}_{n}=\frac{1}{2}\left(\eta_{n,L}+\eta_{n,K}\right)$.

\subsubsection{Thermodynamic equilibrium \label{sec:Thermodynamic-equilibrium}}

In the thermodynamic equilibrium (thermal equilibrium $T_{K}=T_{L}$
and chemical equilibrium $\varphi_{n,K}=\varphi_{n,L}$), both the
discrete current densities (\ref{eq: log mean temp Chatard scheme})
and (\ref{eq: modified drift scheme}) are exactly zero. This is easily
seen from Eqs.~(\ref{eq: modified thermal voltage scheme - discrete thermodynamic form})--(\ref{eq: modified drift scheme - discrete thermodynamic form}),
where the discrete driving force $\left(\varphi_{n,L}-\varphi_{n,K}+P_{n,K,L}^{g/\gamma}\left(T_{L}-T_{K}\right)\right)$
vanishes under thermodynamic equilibrium conditions.

\subsubsection{Strong electric field (drift-dominated limit) \label{sec:Strong-electric-field}}

Due to the asymptotics of the Bernoulli function $B\left(x\to\infty\right)=0$
and $B\left(x\to-\infty\right)\sim-x$, the modified thermal voltage
scheme (\ref{eq: log mean temp Chatard scheme}) approaches the first-order
upwind scheme
\begin{align}
J_{n,K,L}^{g}\left(\delta\phi_{K,L}\to\pm\infty\right) & \sim-qM_{n}\left(\frac{n_{L}+n_{K}}{2}+\frac{n_{K}-n_{L}}{2}\mathop{\mathrm{sign}}\left(\delta\phi_{K,L}\right)\right)\,\delta\phi_{K,L}=J_{n,K,L}^{\text{{upw}}}\label{eq: upwind scheme-1}
\end{align}
in the limit of a strong electrostatic potential gradient $\delta\phi_{K,L}=\phi_{L}-\phi_{K}\to\pm\infty$.
The upwind scheme is a stable, first-order accurate discretization
for advection-dominated problems, where the coefficient is evaluated
in the ``donor cell'' of the flow \citep{Versteeg2007}. Hence,
this asymptotic feature of the original Scharfetter--Gummel scheme,
which is important for the robustness of the discretization as it
avoid spurious oscillations, is preserved in the degenerate and non-isothermal
case. The modified drift scheme (\ref{eq: modified drift scheme})
approaches the upwind scheme as well
\begin{align*}
J_{n,K,L}^{\gamma}\left(\delta\phi_{K,L}\to\pm\infty\right) & \sim-qM_{n}\left(\frac{n_{L}+n_{K}}{2}+\frac{n_{K}-n_{L}}{2}\mathop{\mathrm{sign}}\left(\delta\phi_{K,L}\right)\right)\left(\delta\phi_{K,L}+\frac{k_{B}}{q}\left(\log\left(\frac{\left[\gamma\left(\eta_{n,L}\right)\right]^{T_{L}}}{\left[\gamma\left(\eta_{n,K}\right)\right]^{T_{K}}}\right)-\left(T_{L}-T_{K}\right)\rho_{n,K,L}\right)\right),
\end{align*}
however, in the case of strong degeneration the convergence is significantly
slowed down if the nodal correction factors $\gamma\left(\eta_{n,K}\right)\neq\gamma\left(\eta_{n,K}\right)$
and temperatures $T_{L}\neq T_{K}$ are very different. This is shown
in Fig.~\ref{fig: SG comparison 1D plots}\,(c), where the modified
drift scheme shows a constant offset from the numerically exact solution
of the BVP (\ref{eq: two-point boundary value problem}) for $\delta\phi_{K,L}\to-\infty$.

\subsubsection{No electric field (diffusive limit)}

In the case of a vanishing electrostatic potential gradient $\delta\phi_{K,L}=\phi_{L}-\phi_{K}=0$
the schemes take the form
\begin{align}
\lim_{\delta\phi_{K,L}\to0}J_{n,K,L}^{g} & =M_{n}k_{B}T_{K,L}g_{n,K,L}\left(n_{L}-n_{K}\right),\label{eq: central finite difference}\\
\lim_{\delta\phi_{K,L}\to0}J_{n,K,L}^{\gamma} & =M_{n}k_{B}T_{K,L}\frac{1}{\Lambda\left(\frac{T_{L}^{\rho_{n,K,L}}}{\left[\gamma\left(\eta_{n,L}\right)\right]^{T_{L}/T_{K,L}}},\frac{T_{K}^{\rho_{n,K,L}}}{\left[\gamma\left(\eta_{n,K}\right)\right]^{T_{K}/T_{K,L}}}\right)}\left(\frac{T_{L}^{\rho_{n,K,L}}}{\left[\gamma\left(\eta_{n,L}\right)\right]^{T_{L}/T_{K,L}}}n_{L}-\frac{T_{K}^{\rho_{n,K,L}}}{\left[\gamma\left(\eta_{n,K}\right)\right]^{T_{K}/T_{K,L}}}n_{K}\right).\nonumber 
\end{align}
The modified thermal voltage scheme (\ref{eq: log mean temp Chatard scheme})
approaches the central finite difference discretization (\ref{eq: central finite difference}),
which is a stable discretization for diffusion-dominated transport
problems \citep{Versteeg2007}. With the edge-averaged degeneracy
factor $g_{n,K,L}$, Eq.~(\ref{eq: central finite difference}) nicely
reflects the structure of the diffusive part of the continuous current
density expression (\ref{eq: drift-diffusion currents}) involving
the generalized Einstein relation (\ref{eq: generalized Einstein relations-1}).
For the modified drift scheme (\ref{eq: modified drift scheme}),
the limiting expression is a weighted finite difference discretization.
Due to the different treatment of the degeneracy of the electron gas
via the correction factors (\ref{eq: correction factor-1-1}), it
does not yield a discrete analogue of the generalized Einstein relation.

\subsubsection{Purely thermally driven currents}

In the chemical equilibrium ($\varphi_{\ensuremath{n,L}}=\varphi_{n,K}$),
the current is driven only by the temperature gradient. The corresponding
expressions are easily obtained from Eqs.~(\ref{eq: modified thermal voltage scheme - discrete thermodynamic form})--(\ref{eq: modified drift scheme - discrete thermodynamic form}),
which include the discrete Seebeck coefficients (\ref{eq: edge averaged Seebeck coefficients}).

\subsubsection{Non-negativity of the discrete dissipation rate \label{sec: Non-negativity-of-the discrete dissipation rate}}

The continuous entropy production rate (dissipation rate) per volume
(see Eq.~(\ref{eq: entropy production rate-2}))
\begin{align*}
\dot{s}_{\text{tot}} & =\frac{1}{T}\left(\mu_{c}-\mu_{v}\right)R+\frac{\kappa}{T^{2}}\left\Vert \nabla T\right\Vert ^{2}+\frac{1}{T}H_{J},
\end{align*}
has contributions from carrier recombination, heat flux and Joule
heating $H_{J}=-\left(\nabla\varphi_{n}+P_{n}\nabla T\right)\cdot\mathbf{j}_{n}-\left(\nabla\varphi_{p}+P_{p}\nabla T\right)\cdot\mathbf{j}_{p}$.
With the current density expressions (\ref{eq: current densities-2})
and a recombination rate of the form \eqref{eq: recombination rate}, all
terms in $\dot{s}_{\text{tot}}$ (including $H_{J}$) are evidently
non-negative (i.\,e., zero in the thermodynamic equilibrium and positive
else). Therefore, the model obeys the second law of thermodynamics.
In order to rule out unphysical phenomena such as steady state dissipation \citep{Bessemoulin-Chatard2012,Bessemoulin-Chatard2017},
it is highly desirable to preserve this important structural property
of the continuous system in its discrete counterpart. Given the finite
volume discretization described above, this is straightforwardly achieved
for the contributions from the carrier recombination and the heat
flux, however, it is less obvious for the Joule heating term. In fact,
the non-negativity of the discrete Joule heating term is non-trivial
and can be violated in general when using a naive discretization approach
as in Ref.~\citep{Kato1994}.

We show that the discrete dissipation rate is evidently non-negative
for both generalized Scharfetter--Gummel schemes (\ref{eq: log mean temp Chatard scheme})
and (\ref{eq: modified drift scheme}). This follows immediately
from their consistency with the thermodynamic equilibrium (see Sec.~\ref{sec:Thermodynamic-equilibrium})
in conjunction with the discrete form (\ref{eq: discrete Joule heating})
of the heating term. Substituting Eq.~(\ref{eq: modified thermal voltage scheme - discrete thermodynamic form})
in (\ref{eq: discrete Joule heating}), one obtains
\begin{align*}
H_{J,n}^{g} & =-\left(\varphi_{n,L}-\varphi_{n,K}+P_{n,K,L}^{g}\left(T_{L}-T_{K}\right)\right)J_{n,K,L}^{g}=\sigma_{n,K,L}^{g}\left|\varphi_{n,L}-\varphi_{n,K}+P_{n,K,L}^{g}\left(T_{L}-T_{K}\right)\right|^{2}\geq0,
\end{align*}
which is zero only in the (discrete) thermodynamic equilibrium and
positive else. The discrete conductivity $\sigma_{n,K,L}^{g}$ is
positive by construction, see Eq.~(\ref{eq: modified thermal voltage scheme - discrete thermodynamic form}).
Analogous expressions are obtained for the holes' current contribution
and the modified drift scheme \eqref{eq: modified drift scheme}. In conclusion, the consistency of the
discrete system (\ref{eq: discrete energy-drift-diffusion system})
with the second law of thermodynamics relies on using the respective
Seebeck coefficients $P_{n/p,K,L}$ implied by the current discretization
(see Eq.~(\ref{eq: edge averaged Seebeck coefficients})) consistently
in the discretized Joule heating term (\ref{eq: discrete Joule heating}).
Only then this structural property of the discrete system holds without
any smallness assumption.

\subsection{Comparison with numerically exact solution \label{sec: Comparison with numerically exact solution}}

We investigate the accuracy of the schemes (\ref{eq: log mean temp Chatard scheme})
and (\ref{eq: modified drift scheme}) by comparing them with the
numerically exact solution of the BVP (\ref{eq: two-point boundary value problem}).
In the isothermal case, this has been carried out before in a similar
way by Farrell et al. \citep{Farrell2017a} (for different Scharfetter--Gummel
schemes), which inspired the investigation of highly accurate Scharfetter--Gummel
type discretizations based on the direct numerical integration of
the arising integral equation using quadrature rules in Ref.~\citep{Patriarca2019}.
In the present non-isothermal case, the problem is more complicated
because of the spatially varying temperature distribution along the
edge. It is convenient to recast the problem (\ref{eq: two-point boundary value problem})
into the form
\begin{align}
\begin{aligned}\frac{\mathrm{d}y\left(x\right)}{\mathrm{d}x} & =\frac{\bar{T}}{T\left(x\right)}\left(\delta\Phi+\frac{N_{c}(\bar{T})}{N_{c}(T(x))}\frac{J}{\mathscr{F}\left(y\right)}-\delta\Theta\frac{T(x)N_{c}^{\prime}(T(x))}{N_{c}(T(x))}\frac{\mathscr{F}\left(y\right)}{\mathscr{F}^{\prime}\left(y\right)}\right),\\
y\left(0\right) & =\bar{\eta}_{n}-\frac{1}{2}\delta\eta_{n},\qquad\qquad y\left(1\right)=\bar{\eta}_{n}+\frac{1}{2}\delta\eta_{n},
\end{aligned}
\label{eq: two-point boundary value problem - eta form}
\end{align}
with the notations $T\left(x\right)\equiv\left(1+\left[x-\frac{1}{2}\right]\delta\Theta\right)\bar{T}$,
$\bar{T}=\frac{1}{2}\left(T_{L}+T_{K}\right)$, $\delta T=T_{L}-T_{K}$,
$\bar{\eta}_{n}=\frac{1}{2}\left(\eta_{n,L}+\eta_{n,K}\right)$, $\delta\eta_{n}=\eta_{n,L}-\eta_{n,K}$
and the non-dimensionalized quantities
\begin{align*}
J & =\frac{J_{n,K,L}}{M_{n}k_{B}\bar{T}N_{c}(\bar{T})}, & \delta\Phi & =\frac{q\left(\phi_{L}-\phi_{K}\right)}{k_{B}\bar{T}}, & \delta\Theta & =\frac{\delta T}{\bar{T}}.
\end{align*}
The exact current $J_{\text{exact}}=J_{\text{exact}}(\delta\Phi,\delta\eta_{n},\delta\Theta,\bar{\eta}_{n},\bar{T})$
is a function of five parameters that satisfies the BVP (\ref{eq: two-point boundary value problem - eta form}).
We solve the BVP \eqref{eq: two-point boundary value problem - eta form}
numerically using the shooting method \citep{Keller1976}, where we
combine a 4th order Runge--Kutta method with Brent's root finding
algorithm \citep{Brent1971} . The problem is invariant under
the simultaneous transformation
\begin{align*}
\delta\Phi & \to-\delta\Phi, & \delta\eta_{n} & \to-\delta\eta_{n}, & \delta\Theta & \to-\delta\Theta, & x & \to1-x, & J & \to-J
\end{align*}
(i.\,e., the sign of the current changes when changing the nodes
$K\leftrightarrow L$), such that we can restrict our analysis to
$\delta\Theta\geq0$, when exploring the accuracy of the discrete
current in the $\left(\delta\Phi,\delta\eta_{n}\right)$-plane. The
comparison is carried out for $\mathscr{F}\left(\eta\right)=F_{1/2}\left(\eta\right)$
and $N_{c}=2\left(m_{c}^{\ast}k_{B}T/(2\pi\hbar^{2})\right)^{3/2}$.
The Fermi--Dirac integrals $F_{1/2}\left(\eta\right)$ and $F_{-1/2}\left(\eta\right)$
are evaluated using MacLeod's algorithm \citep{MacLeod1998}.

\begin{figure}
\centering

\includegraphics[width=1\columnwidth]{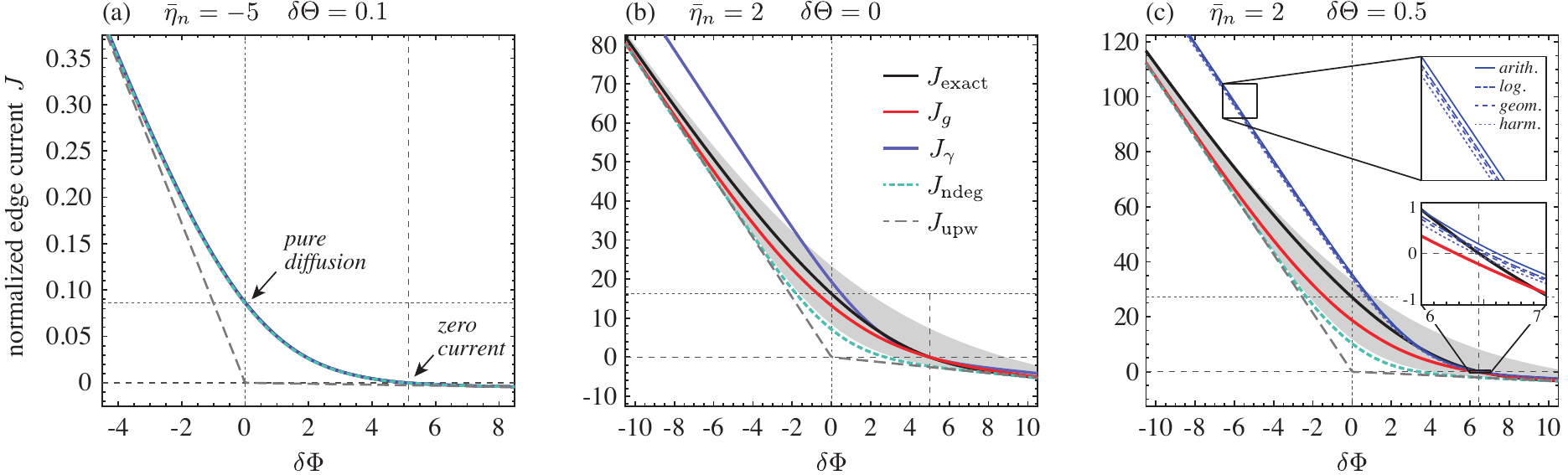}

\caption{Comparison of the non-isothermal Scharfetter--Gummel schemes for
$\delta\eta_{n}=5$. (a)~In the non-degenerate regime ($\bar{\eta}_{n}=-5$)
all schemes coincide, even in the presence of a temperature gradient.
(b)~At $\bar{\eta}_{n}=2$ degeneration effects become significant.
While $J_{g}$ follows $J_{\text{exact }}$ with an acceptable error
over the whole range of $\delta\Phi$, the modified drift scheme $J_{\gamma}$
has a significant offset at strong electric fields ($\delta\Phi\to-\infty$).
The grey shaded area indicates the analytic bounds of the modified
thermal voltage scheme determined using the nodal values $g_{n,K}$
and $g_{n,L}$ instead of $g_{n,K,L}$. (c)~An additional temperature
gradient increases the error in the degenerate regime. The insets
show the effect of different averages $\rho_{n,K,L}$ of the nodal
values $\rho_{n,K}$ and $\rho_{n,L}$ in the modified drift scheme
and the different behavior of the schemes in the region of vanishing
discrete currents.}

\label{fig: SG comparison 1D plots}
\end{figure}

Figure \ref{fig: SG comparison 1D plots} shows the numerically exact
current $J_{\text{exact}}$ along with the approximations $J_{g}$
(modified thermal voltage scheme (\ref{eq: log mean temp Chatard scheme}))
and $J_{\gamma}$ (modified drift scheme (\ref{eq: modified drift scheme}))
as a function of the normalized electric field $\delta\Phi$ along
the edge. For weak degeneracy, both schemes agree with the numerically
exact solution -- even in the case of non-isothermal conditions.
This is shown in Fig.~\ref{fig: SG comparison 1D plots}\,(a) for
$\bar{\eta}_{n}=-5$ and $\delta\Theta=0.1$. At strong electric fields
$\delta\Phi\to\pm\infty$, all schemes approach the upwind scheme
$J_{\text{upw}}$ (grey dashed line, cf. Eq.~(\ref{eq: upwind scheme-1})).
The schemes (\ref{eq: log mean temp Chatard scheme}) and (\ref{eq: modified drift scheme})
differ in the treatment of degeneration effects, which becomes apparent
for increased $\bar{\eta}_{n}$. Figure~\ref{fig: SG comparison 1D plots}\,(b)
shows the results for $\bar{\eta}_{n}=2$ at isothermal conditions
$\delta\Theta=0$. The modified thermal voltage scheme $J_{g}$ (red
line) yields an acceptable deviation from the exact result $J_{\text{exact}}$
(black line) over the whole range of $\delta\Phi$. The error vanishes
at strong electric fields where both $J_{g}$ and $J_{\text{exact}}$
converge to the upwind scheme $J_{\text{upw}}$. The modified drift
scheme (purple line), however, shows a significant error at large
(negative) $\delta\Phi$, where it overestimates the current density
significantly (about 33~\% relative error at $\delta\Phi=-3$). This
behavior results from the different treatment of the degeneration
effects, that degrades the convergence of the modified drift scheme
in the case of strong degeneration (see Sec.~\ref{sec:Strong-electric-field}).
The plot highlights two other important exceptional points (pure diffusion
and zero current), where both schemes show a similar accuracy. In
the presence of an additional temperature gradient along the edge,
see Fig.~\ref{fig: SG comparison 1D plots}\,(c), the approximation
error of both schemes increases. The upper inset shows that the choice
of the average of $\rho_{n,K,L}$ (see Eq.~(\ref{eq: deviation function}))
has only a minor impact on the modified drift scheme. The lower inset
zooms on the region where the currents become zero. Here, all schemes
provide a satisfying accuracy, but none of them is exact, i.\,e.,
they yield a small spurious discrete current and intersect with the
exact solution only in the vicinity of the exact zero current point.
We observe that the modified drift schemes show a slightly better
performance in this case, i.\,e., the Seebeck coefficient (\ref{eq: edge averaged Seebeck coefficients - correction factor})
appears to be slightly better than (\ref{eq: edge averaged Seebeck coefficients - mod thermal voltage}).

\begin{figure}
\centering

\includegraphics[width=1\textwidth]{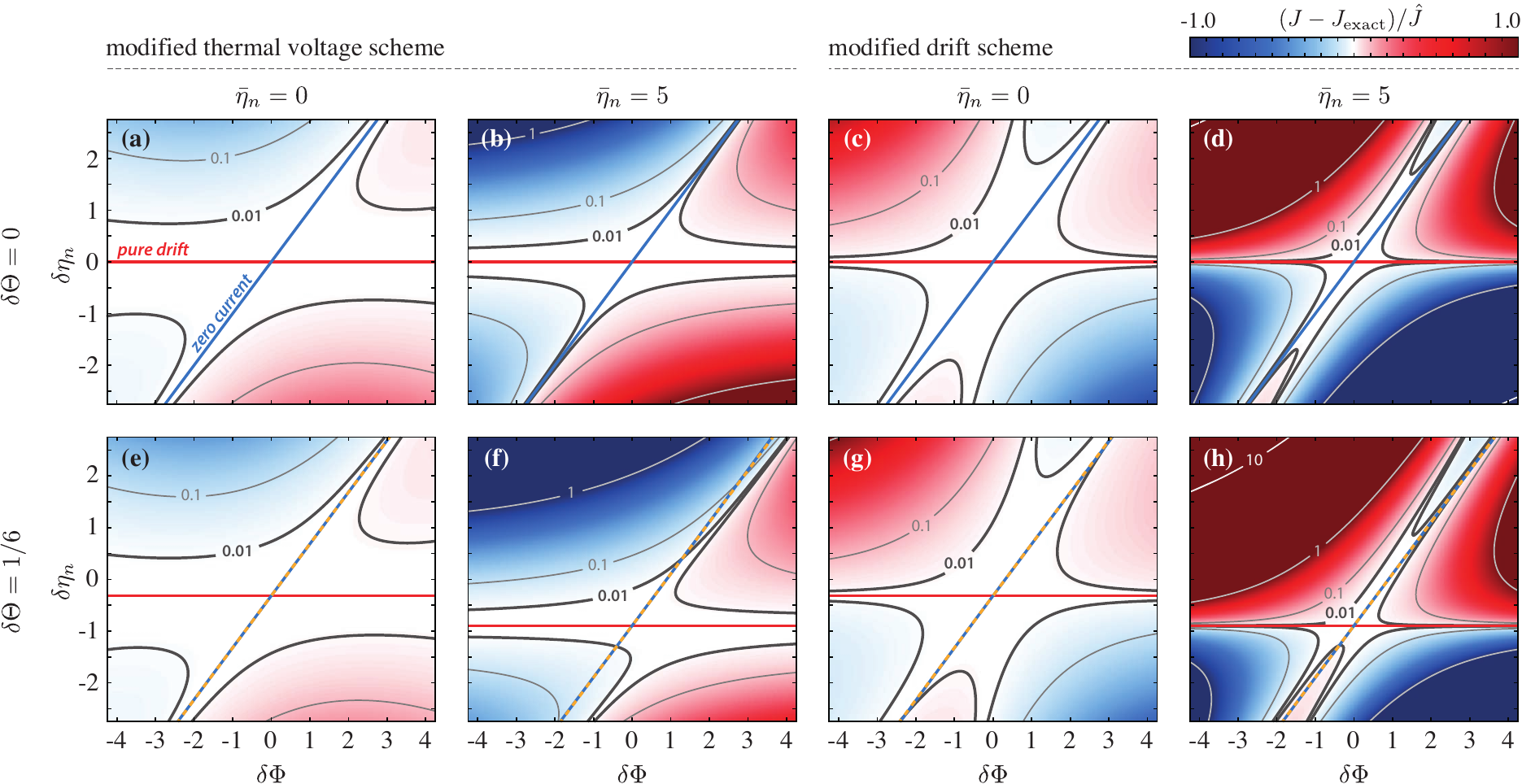}

\caption{Comparison of the two non-isothermal Scharfetter--Gummel schemes
(\ref{eq: log mean temp Chatard scheme}) and (\ref{eq: modified drift scheme})
(with arithmetic average $\rho_{n,K,L}=\left(\rho_{n,K}+\rho_{n,L}\right)/2$)
in the ($\delta\Phi,\delta\eta_{n}$)-plane at isothermal ($\delta\Theta=0$,
top row (a)--(d)) and non-isothermal ($\delta\Theta=1/6$, bottom
row (e)--(h)) conditions and different levels of degeneration (weak
$\bar{\eta}_{n}=0$ or strong $\bar{\eta}_{n}=5$). The normalized
absolute error $(J-J_{\text{exact}})/\hat{J}$ is color-coded for
the range $[-1,1]$ (dark colored regions correspond to larger errors,
see the level lines). See the text for a discussion.}
\label{fig: SG comparison 2D plots}
\end{figure}

The normalized absolute errors $(J-J_{\text{exact}})/\hat{J}$ (with
$\hat{J}=M_{n}k_{B}\bar{T}N_{c}(\bar{T})$) of the two schemes (\ref{eq: log mean temp Chatard scheme})
and (\ref{eq: modified drift scheme}) are shown in the ($\delta\Phi,\delta\eta_{n}$)-plane
in Fig.~\ref{fig: SG comparison 2D plots} under isothermal ($\delta\Theta=0$,
top row (a)--(d)) and non-isothermal ($\delta\Theta=1/6$, bottom
row (e)--(h)) conditions and for different levels of degeneration
(weak $\bar{\eta}_{n}=0$ and strong $\bar{\eta}_{n}=5$). In the
limit of very fine meshes $(\delta\Phi,\delta\eta_{n},\delta\Theta)\to(0,0,0)$,
both schemes coincide and the deviation from the exact current approaches
zero. One observes that the size of the white regions with a normalized
absolute error below $0.01$ (in the following denoted as ``low error
domain'') are generally larger for the modified thermal voltage scheme
than for the modified drift scheme. Thus, the modified thermal voltage
scheme is expected to yield a higher accuracy on sufficiently fine
meshes. This will be evidenced by the numerical simulation of a p-n-diode
in Sec.~\ref{sec: benchmark simulation}. The plots in Fig.~\ref{fig: SG comparison 2D plots}
feature two additional lines, that refer to special limiting cases
where both schemes yield a very high accuracy. The case of a pure
drift current (i.\,e., no diffusion $n_{L}=n_{K}$) is indicated
by a red line; the zero current line (blue) refers to the curve in
the ($\delta\Phi,\delta\eta_{n}$)-plane where the exact current vanishes
($J_{\text{exact}}=0$ in the BVP (\ref{eq: two-point boundary value problem - eta form})).
In the isothermal case ($\delta\Theta=0$), the latter corresponds
to the thermodynamic equilibrium, in the non-isothermal case $(\delta\Theta\neq0)$
it refers to the situation of compensating chemical and thermal driving
forces. Both schemes are exact in the case of a pure drift current,
i.\,e., they asymptotically approach the upwind scheme (\ref{eq: upwind scheme-1}),
which is important for the robustness of the discretization in order
to avoid spurious oscillations. The modified thermal voltage scheme
shows a high accuracy also for slight deviations from the pure drift
line, even in the case of strong degeneration, see Fig.~\ref{fig: SG comparison 2D plots}\,(b,\,f).
In contrast, the modified drift scheme yields significant errors (much
higher than $0.01$) already for tiny deviations from the pure drift
line in the strongly degenerate case, see Fig.~\ref{fig: SG comparison 2D plots}\,(d,\,h).
This behavior has already been observed above in Fig.~\ref{fig: SG comparison 1D plots}\,(b,\,c)
and was predicted analytically in Sec.~\ref{sec:Strong-electric-field}.
Note that in the non-isothermal case, the temperature gradient shifts
the pure drift line from $\delta\eta_{n}^{\text{pure drift}}\vert_{\delta\Theta=0}=0$
(in Fig.~\ref{fig: SG comparison 2D plots}\,(a--d)) to  $\delta\eta_{n}^{\text{pure drift}}\approx-\delta\Theta g_{\eta}(\bar{\eta}_{n})\bar{T}N_{c}^{\prime}(\bar{T})/N_{c}(\bar{T})$
(see Fig.~\ref{fig: SG comparison 2D plots}\,$\mbox{(e\textendash h)}$).
A prominent feature of the modified drift scheme is the additional
intersection with the exact solution (see also Fig.~\ref{fig: SG comparison 1D plots}\,(c)
at $\delta\Phi\approx3.5$), which leads to additional ``fingers''
of the low error domain, see Fig.~\ref{fig: SG comparison 2D plots}\,(c,\,d,\,g,\,h),
that are not associated with any special limiting case. The same feature
has been observed for the so-called \emph{inverse activity scheme}
described in Ref.~\citep{Farrell2017a}. Finally, we study the consistency
of the discretization schemes with the zero current line. In the isothermal
case, both schemes are exact and therefore consistent with the thermodynamic
equilibrium, see Fig.~\ref{fig: SG comparison 2D plots}\,(a--d).
In the strongly degenerate case, however, the zero current line is
only partially located within the low error domain, since the schemes
intersect with the exact solution only in the vicinity of the zero
current line and not exactly on it (see also the inset of Fig.~\ref{fig: SG comparison 1D plots}\,(c)).
Nevertheless, the zero current lines of the discrete schemes, which
are plotted as dashed orange lines in Fig.~\ref{fig: SG comparison 2D plots}\,(e--h),
nicely overlap with the exact zero current line (blue). Thus, the
spurious non-zero currents are very low and the little discrepancy
in this limiting case is only of minor importance.

\subsection{Analytical error estimate \label{sec: Analytical error estimate}}

We compare both schemes (\ref{eq: log mean temp Chatard scheme})
and (\ref{eq: modified drift scheme}) by deriving an upper error
bound. We follow the approach developed by Farrell et al. \citep{Farrell2017a}
and extend it to the non-isothermal case. Using the identities
\begin{align*}
B\left(x\right)-B\left(-x\right) & =-x, & B\left(x\right)+B\left(-x\right) & =x\coth{\left(\frac{x}{2}\right)},
\end{align*}
we obtain the series expansion of the discrete currents (\ref{eq: log mean temp Chatard scheme})
and (\ref{eq: modified drift scheme}) at $\left(\delta\Phi,\delta\eta_{n},\delta\Theta\right)=\left(0,0,0\right)$
up to second order as
\begin{align*}
J_{g} & =-\mathscr{F}\left(\bar{\eta}_{n}\right)\delta\Phi+\mathscr{F}\left(\bar{\eta}_{n}\right)\left(\delta\eta+\frac{\bar{T}N_{c}^{\prime}(\bar{T})}{N_{c}(\bar{T})}g_{\eta}\left(\bar{\eta}_{n}\right)\delta\Theta\right)\frac{\tilde{X}_{g}}{2}\coth{\left(\frac{\tilde{X}_{g}}{2}\right)}+\mathcal{O}(\delta^{3}),\\
J_{\gamma} & =-\mathscr{F}\left(\bar{\eta}_{n}\right)\tilde{X}_{\gamma}+\mathscr{F}^{\prime}\left(\bar{\eta}_{n}\right)\left(\delta\eta+\frac{\bar{T}N_{c}^{\prime}(\bar{T})}{N_{c}(\bar{T})}g_{\eta}\left(\bar{\eta}_{n}\right)\delta\Theta\right)\frac{\tilde{X}_{\gamma}}{2}\coth{\left(\frac{\tilde{X}_{\gamma}}{2}\right)}+\mathcal{O}(\delta^{3}),
\end{align*}
where
\begin{align*}
\tilde{X}_{g} & =\frac{1}{g_{\eta}\left(\bar{\eta}_{n}\right)}\delta\Phi, & \tilde{X}_{\gamma} & =\delta\Phi-\frac{g_{\eta}\left(\bar{\eta}_{n}\right)-1}{g_{\eta}\left(\bar{\eta}_{n}\right)}\left(\delta\eta+\frac{\bar{T}N_{c}^{\prime}(\bar{T})}{N_{c}(\bar{T})}g_{\eta}\left(\bar{\eta}_{n}\right)\delta\Theta\right),
\end{align*}
and $\mathcal{O}(\delta^{3})\equiv\mathcal{O}(\delta\eta_{n}^{3})+\mathcal{O}(\delta\eta_{n}^{2}\,\delta\Theta)+\mathcal{O}(\delta\Theta^{2}\,\delta\eta_{n})+\mathcal{O}(\delta\Theta^{3})+\mathcal{O}(\delta\Phi\,\delta\eta_{n}^{2})+\mathcal{O}(\delta\Phi\,\delta\eta_{n}\,\delta\Theta)+\mathcal{O}(\delta\Phi\,\delta\Theta^{2})$
denotes the third-order corrections. The second-order expansion of
the modified drift scheme is independent of the kind of average used
for $\rho_{n,K,L}\approx\rho_{n}(\bar{\eta}_{n},\bar{T})+\mathcal{O}(\delta\eta_{n}^{2})+\mathcal{O}(\delta\eta_{n}\,\delta\Theta)+\mathcal{O}(\delta\Theta^{2})$,
as only its zeroth-order contribution (where all means coincide) is
relevant here. Using the inequality \citep{Farrell2017a}
\[
1\leq x\coth{\left(x\right)}\leq1+\left|x\right|,
\]
we arrive at the error estimates for the modified thermal voltage
scheme (neglecting third-order terms)
\begin{align}
\left|J_{g}-J_{1}\right| & \leq\frac{1}{2}\mathscr{F}^{\prime}\left(\bar{\eta}_{n}\right)\left(\left|\delta\Phi\,\delta\eta_{n}\right|+\frac{\bar{T}N_{c}^{\prime}(\bar{T})}{N_{c}(\bar{T})}g_{\eta}\left(\bar{\eta}_{n}\right)\left|\delta\Phi\,\delta\Theta\right|\right)\label{eq: Chatard error bound}
\end{align}
and the modified drift scheme
\begin{align}
\left|J_{\gamma}-J_{1}\right| & \leq\frac{1}{2}\mathscr{F}^{\prime}\left(\bar{\eta}_{n}\right)\left(\left|\delta\Phi\,\delta\eta_{n}\right|+\frac{\bar{T}N_{c}^{\prime}(\bar{T})}{N_{c}(\bar{T})}g_{\eta}\left(\bar{\eta}_{n}\right)\left|\delta\Phi\,\delta\Theta\right|+\frac{g_{\eta}\left(\bar{\eta}_{n}\right)-1}{g_{\eta}\left(\bar{\eta}_{n}\right)}\left|\delta\eta_{n}+\frac{\bar{T}N_{c}^{\prime}(\bar{T})}{N_{c}(\bar{T})}g_{\eta}\left(\bar{\eta}_{n}\right)\delta\Theta\right|^{2}\right),\label{eq: correction factor error bound}
\end{align}
where $J_{1}=\mathscr{F}\left(\bar{\eta}_{n}\right)\left(\delta\eta_{n}-\delta\Phi+\frac{\bar{T}N_{c}^{\prime}(\bar{T})}{N_{c}(\bar{T})}g_{\eta}\left(\bar{\eta}_{n}\right)\delta\Theta\right)$
is the first-order exact solution of the BVP (\ref{eq: two-point boundary value problem - eta form}).
Both schemes converge to the exact result as their first-order terms
agree with $J_{1}$. The first two terms of Eq.~(\ref{eq: correction factor error bound})
coincide with Eq.~(\ref{eq: Chatard error bound}). The error bound
for the modified drift scheme has an additional second-order contribution
that becomes significant in the case of strong degeneration $\bar{\eta}\gg1$
where $(g_{\eta}\left(\bar{\eta}_{n}\right)-1)/g_{\eta}\left(\bar{\eta}_{n}\right)\to1$.
Therefore, the maximum error of the modified thermal voltage scheme
is guaranteed to be smaller than that of the modified drift scheme
in the case of degenerate carrier statistics. This analytical result
is consistent with the numerical results shown in Figs.~\ref{fig: SG comparison 1D plots}
and \ref{fig: SG comparison 2D plots} and holds in both the isothermal
and the non-isothermal case. For non-degenerate carrier statistics,
both error estimates (\ref{eq: Chatard error bound}) and (\ref{eq: correction factor error bound})
coincide, since both schemes reduce to the non-degenerate scheme (\ref{eq: non-degenerate, non-isothermal SG}).

\subsection{Numerical simulation of a p-n-diode \label{sec: benchmark simulation}}

\begin{figure}
\includegraphics[width=1\textwidth]{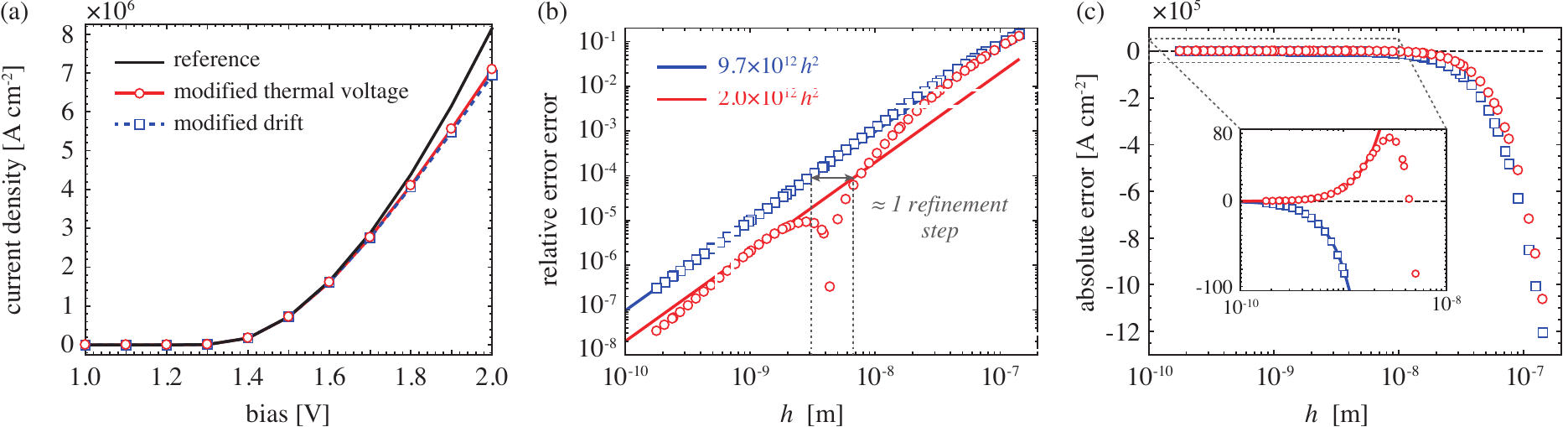}\caption{Convergence of total current density in the p-n diode problem without
self-heating effects ($H=0$, isothermal case). (a)~Current-voltage
curves obtained by the two schemes (\ref{eq: log mean temp Chatard scheme})
and (\ref{eq: modified drift scheme}) on an equidistant grid with
13 nodes. The reference solution (black line) was obtained on a fine
grid (65535 nodes). (b)~Convergence of the relative error with respect
to the reference solution under mesh refinement at $2\,\text{V}$.
The modified drift scheme (blue squares) shows a monotonous, quadratic
convergence for decreasing $h$. The modified thermal voltage scheme
(red circles) converges non-monotonously as it intersects with the
reference solution at $h\approx4.5\,\text{nm}$. (c)~Convergence
of the absolute error of both schemes.}
\label{fig: convergence isothermal}
\end{figure}

We consider a one-dimensional GaAs-based p-n-diode and compare the
convergence of the total current density ($\mathbf{j}_{\text{tot}}=\mathbf{j}_{n}+\mathbf{j}_{p}$)
under mesh refinement using both discretization schemes. The device
consists of a $1\,\text{\textmu m}$ n-doped section with $C=N_{D}^{+}=2\times10^{18}\,\text{cm}^{-3}$
followed by a $1\,\text{\textmu m}$ long p-doped section with $C=-N_{A}^{-}=-2\times10^{18}\,\text{cm}^{-3}$.
We use Fermi--Dirac statistics $\mathscr{F}\left(\eta\right)=F_{1/2}\left(\eta\right)$
and take Shockley--Read--Hall recombination, spontaneous emission
and Auger recombination into account \citep{Selberherr1984,Palankovski2004}.
The material parameters, mobility models (depending on temperature
and doping density) and the temperature-dependent heat conductivity
model are taken from Ref.~\citep{Palankovski2004}. The mobilities
and thermal conductivity along the edges are taken as the harmonic
average of the respective nodal values. We use Dirichlet boundary
conditions on both ends of the diode, modeling ideal Ohmic contacts
(charge neutrality at the boundary) and ideal heat sinks with $T_{\text{contact}}=300\,\text{K}$.
The simulations are carried out on equidistant grids with varying
number of mesh points $N_{\text{nodes}}$ and mesh size $h=2\,\text{\textmu m}/(N_{\text{nodes}}-1)$.
The nonlinear systems are solved using a Newton iteration method with a fully analytical Jacobian matrix \cite{Farrell2017}.

Figure~\ref{fig: convergence isothermal}\,(a) shows the current-voltage
curves obtained by both discretization schemes on a coarse grid (13
nodes, $h\approx1.7\times10^{-7}\,\text{m}$) under isothermal conditions,
i.\,e., without self-heating and Seebeck effect. For the evaluation
of the error, we use a reference solution that was computed on a fine
grid with 65535 nodes ($h\approx3.1\times10^{-11}\,\text{m}$), where
the relative error between both schemes is about $9.6\times10^{-9}$.
At $2\thinspace\text{V}$ the computed currents differ significantly
from the reference result: The relative error is about $13\,\%$ for
the modified thermal voltage scheme and $15\,\%$ for the modified
drift scheme. The convergence of the computed current densities to
the reference result under mesh refinement is shown in Fig.~\ref{fig: convergence isothermal}\,(b).
The modified drift scheme (Eq.~(\ref{eq: modified drift scheme}),
blue squares) shows a monotonous, quadratic convergence for decreasing
$h$. The modified thermal voltage scheme (Eq.~(\ref{eq: log mean temp Chatard scheme}),
red circles), however, shows a non-monotonous convergence behavior
as it intersects with the reference solution at $h\approx4.5\times10^{-9}\,\text{m}$.
On sufficiently fine meshes ($h<10^{-8}\,\text{m}$), the error of
the modified thermal voltage scheme is almost one order of magnitude
smaller than that of the modified drift scheme. Conversely, the modified
thermal voltage scheme reaches the same accuracy as the modified drift
scheme already on a coarse grid with less than half of the number
of nodes. Thus, the modified thermal voltage scheme saves about one
refinement step. The convergence of the absolute error is plotted
in Fig.~\ref{fig: convergence isothermal}\,(c), where the inset
highlights the origin of the non-monotonous convergence behavior of
the modified thermal voltage scheme.

\begin{figure}
\includegraphics[width=1\textwidth]{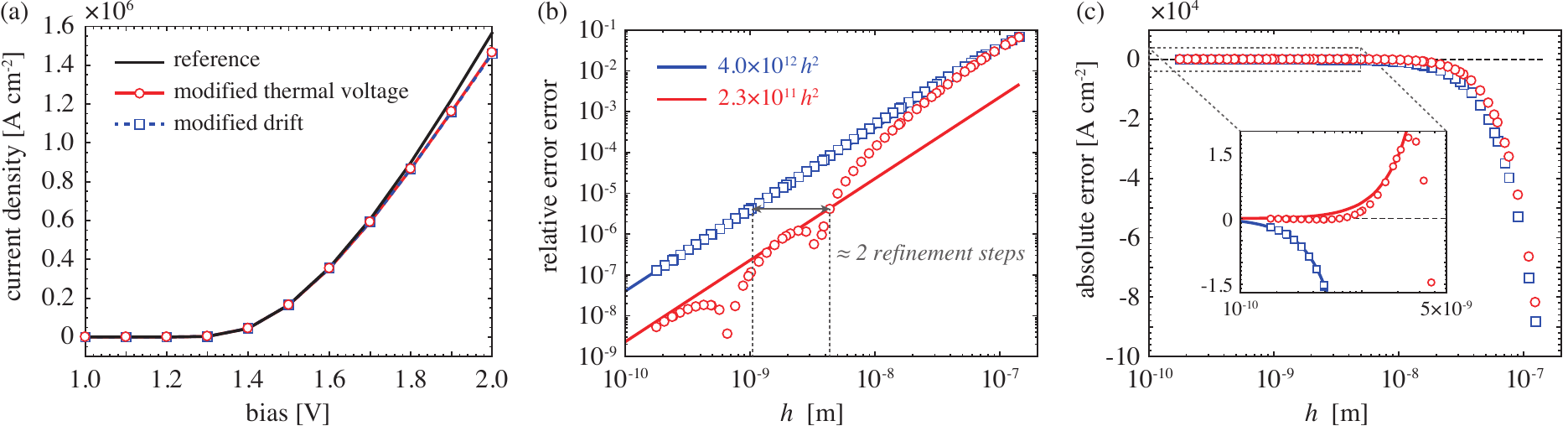}\caption{Convergence of total current density in the p-n diode problem with
self-heating effects. (a)~Current-voltage curves obtained by the
two schemes (\ref{eq: log mean temp Chatard scheme}) and (\ref{eq: modified drift scheme})
on an equidistant grid with 13 nodes. As in Fig.~\ref{fig: convergence isothermal},
the reference solution (black line) was obtained on a grid with 65535
nodes. Self-heating lowers the mobilities such that the total current
density is smaller than in the isothermal case, cf. Fig.~\ref{fig: convergence isothermal}\,(a).
(b)~Convergence of the relative error with respect to the reference
solution under mesh refinement at $2\,\text{V}$. As in the isothermal
case, the modified thermal voltage scheme (red circles) shows a non-monotonous
convergence behavior that is faster than the quadratic convergence
of the modified drift scheme (blue squares). (c)~Convergence of the
absolute error of both schemes. The modified thermal voltage scheme
intersects twice with the reference solution.}
\label{fig: convergence non-isothermal}
\end{figure}

The numerical results for the non-isothermal case, where self-heating
and the Seebeck effect are taken into account, are shown in Fig.~\ref{fig: convergence non-isothermal}.
The results are qualitatively very similar to the isothermal case
shown in Fig.~\ref{fig: convergence isothermal}; quantitatively
the advantage of the modified thermal voltage scheme over the modified
drift scheme is even greater. On a coarse grid, the total current
is underestimated by both schemes with a relative error of about $8\,\%$,
see Fig.~\ref{fig: convergence non-isothermal}\,(a). For sufficiently
fine meshes ($h<0.5\times10^{-8}\text{m}$), the error of the modified
thermal voltage scheme is always more than one order of magnitude
smaller than that of the modified drift scheme, see Fig.~\ref{fig: convergence non-isothermal}\,(b).
In other words, the modified thermal voltage scheme reaches the same
accuracy already on an about four times coarser mesh (two refinement
steps), which is a substantial advantage for large problems involving
complex multi-dimensional geometries. Again, we observe a non-monotonous
convergence behavior of the modified thermal voltage scheme, see Fig.~\ref{fig: convergence non-isothermal}\,(b,\,c).

\section{Summary and conclusion}

We discussed the non-isothermal drift-diffusion system for the simulation
of electro-thermal transport processes in semiconductor devices. It
was shown that the model equations take a remarkably simple form
when assuming the Kelvin formula for the Seebeck coefficient. First,
the heat generation rate involves exactly the three classically known
self-heating effects (Joule heating, recombination heating, Thomson--Peltier
effect) without any further transient contributions. Moreover, our
modeling approach immediately yields the correct average kinetic energy
of the carriers in the recombination heating term, independently of any
scattering parameter. Second, the Kelvin formula enables a simple
representation of the electrical current densities in the drift-diffusion
form, where the thermal driving force can be entirely absorbed in
the (nonlinear) diffusion coefficient via the generalized Einstein
relation. The Kelvin formula accounts for the degeneration of the
electron-hole plasma (Fermi--Dirac statistics) and was shown to be
in a good quantitative agreement with experimental data reported for
n-GaAs. 

We have derived two non-isothermal generalizations of the finite volume
Scharfetter--Gummel scheme for the discretization of the current
densities, which differ in their treatment of degeneration effects.
The first approach is based on an approximation of the discrete generalized
Einstein relation and implies a specific modification of the thermal
voltage. The second scheme is based on including the degeneration
effects into a modification of the electric field, which is similar
to the conventional method that is widely used in commercial device
simulation software packages \citep{Synopsys2010,Silvaco2016}. We
presented a detailed analysis of both schemes by assessing their accuracy
in comparison to the numerically exact solution of the underlying
two-point boundary value problem. Moreover, we derived analytical
error bounds and investigated important structure preserving properties
of the discretizations, including the consistency with the thermodynamic
equilibrium, the non-negativity of the discrete dissipation rate (second
law of thermodynamics on the discrete level) and their asymptotic
behavior in the drift- and diffusion-dominated limits. Finally, we
performed a numerical convergence study for a simple example case. Our results indicate a significantly
higher accuracy and faster convergence of the modified thermal
voltage scheme in comparison to the modified drift scheme. This result
holds under both isothermal and non-isothermal conditions. The
higher accuracy --- about one order of magnitude for sufficiently
fine grids in the present case study --- of the modified thermal
voltage scheme makes it a favorable discretization method for problems
exhibiting stiff solutions (internal layers at p-n junctions, boundary
layers at electrical contacts) or devices with a complicated
multi-dimensional geometry, where the number of nodes required to
reach the asymptotic accuracy regime is extremely large and routinely
exceeds the available computational power.

In more general situations, where the Seebeck coefficient deviates
from the Kelvin formula (e.\,g., due to the phonon drag effect),
we suggest to combine the two discretization techniques by decomposing
the Seebeck coefficient into a Kelvin formula part and an excess contribution:
$P_{n}=P_{n}^{\text{Kelvin}}+P_{n}^{\text{exc}}$. The first part
can be absorbed in the generalized Einstein relation, which allows
for the treatment described in Sec.~\ref{sec: Modified thermal voltage scheme}
and inherits the improved accuracy of the modified thermal voltage
scheme. The excess part $P_{n}^{\text{exc}}$ must be averaged along
the edge and plays a similar role as the $\rho_{n,K,L}$ term in Sec.~\ref{sec: Correction factor scheme}
(leading to an additive correction in the argument of the Bernoulli
function).

\section*{Acknowledgements}

This work was funded by the German Research Foundation (DFG) under
Germany's Excellence Strategy -- EXC2046: \textsc{Math}+ (project
AA2-3). The author is grateful to Thomas Koprucki for carefully reading
the manuscript and giving valuable comments.

\appendix

\renewcommand*{\thesection}{\appendixname~\Alph{section}}

\section{Non-parabolic energy bands} \label{sec: non-parabolic energy dispersion}

The results presented in this paper hold for Fermi--Dirac statistics
with arbitrary density of states, provided that the Kelvin
formula for the Seebeck coefficient is applicable. In particular,
this includes also the case of non-parabolic energy bands. For example,
we assume the scalar Kane-type model \cite{Lundstrom2000} for
the conduction band energy dispersion
\begin{align*}
E & =E_{c}+\varepsilon_{k}, & \left(1+\alpha\varepsilon_{k}\right)\varepsilon_{k} & =\frac{\hbar^{2}k^{2}}{2m_{c}^{\ast}}
\end{align*}
with the \emph{non-parabolicity parameter} $\alpha\geq0.$ This corresponds to the density of states (in 3D)
\begin{equation*}
D_{c}\left(E\right) =\frac{1}{V}\sum_{\mathbf{k},\sigma}\delta\left(E-E\left(\mathbf{k}\right)\right) =\frac{1}{2\pi^{2}}\left(\frac{2m_{c}^{\ast}}{\hbar^{2}}\right)^{3/2}\sqrt{\left(E-E_{c}\right)\left(1+\alpha\left(E-E_{c}\right)\right)}\left(1+2\alpha\left(E-E_{c}\right)\right)\,\Theta\left(E-E_{c}\right),
\end{equation*}
which recovers with the parabolic case for $\alpha=0$. Here, $\Theta$ is the Heaviside step function. 
For weak non-parabolicity $0\leq\alpha/\beta \ll1$, the corresponding electron density
can be expanded in a series of Fermi--Dirac integrals \eqref{eq: Fermi-Dirac integral}
\begin{align*}
n &=\int_{\mathbb{R}}\mathrm{d}E\,D_{c}\left(E\right)\frac{1}{\exp{\left(\beta\left[E-q\phi-\mu_{c}\right]\right)}+1}=N_{c}\left(T\right)\mathscr{F}\left(\eta_{n}\right), & \mathscr{F}\left(\eta_{n}\right) &= \sum_{n=0}^{\infty}\frac{1}{n!}\frac{\Gamma\left(\frac{5}{2}+n\right)}{\Gamma\left(\frac{5}{2}-n\right)}\left(\frac{\alpha}{\beta}\right)^{n}F_{n+\frac{1}{2}}\left(\eta_{n}\right),
\end{align*}
with $\eta_{n}=\beta\left(\mu_{c}+q\phi-E_{c}\right)$ and $\beta=\left(k_{B}T\right)^{-1}$,
which is clearly of the type \eqref{eq: carrier density state equations}.

The desired asymptotic behavior in the non-degenerate limit $\mathscr{F}(\eta \ll -1) \sim \exp{(\eta)}$ is restored by rescaling
\begin{align*}
N_c\left(T\right) &\to  \sqrt{\frac{\beta}{\pi\alpha}}\exp{\left(\frac{\beta}{2\alpha}\right)}K_{2}\left(\frac{\beta}{2\alpha}\right) N_c\left(T\right), & \mathscr{F}(\eta) &\to \left(\sqrt{\frac{\beta}{\pi\alpha}}\exp{\left(\frac{\beta}{2\alpha}\right)}K_{2}\left(\frac{\beta}{2\alpha}\right)\right)^{-1}\mathscr{F}(\eta),
\end{align*}
where $K_\nu$ is the modified Bessel function of second kind.

\section{Generalization to the case of multiple temperatures} \label{sec: Generalization to the case of multiple temperatures}

The results for the Kelvin formula can be generalized
to the multi-temperature case, i.e., to systems describing hot carrier transport
where the carrier ensembles have temperatures different from the lattice temperature $T_{n}\neq T_{L}\neq T_{p}$.
We postulate the free energy density of the electron-hole plasma
\begin{align}
\begin{aligned} \label{eq: free energy density e-h multi-temperature}
f_{\text{e--h}}\left(n,p,T_{n},T_{p},T_{L}\right) & =k_{B}T_{n}\left[n\mathscr{F}^{-1}\left(\frac{n}{N_{c}\left(T_{L},T_{n}\right)}\right)-N_{c}\left(T_{n},T_{L}\right)\mathscr{G}\left(\mathscr{F}^{-1}\left(\frac{n}{N_{c}\left(T_{n},T_{L}\right)}\right)\right)\right]+nE_{c}\left(T_{L}\right)\\
 & \phantom{=}+k_{B}T_{p}\left[p\mathscr{F}^{-1}\left(\frac{p}{N_{v}\left(T_{p},T_{L}\right)}\right)-N_{v}\left(T_{p},T_{L}\right)\mathscr{G}\left(\mathscr{F}^{-1}\left(\frac{p}{N_{v}\left(T_{p},T_{L}\right)}\right)\right)\right]-pE_{v}\left(T_{L}\right),
\end{aligned}
\end{align}
where the effective density of states $N_{c/v}=N_{c/v}(T_{n/p},T_L)$
is now a function of the carrier temperature and the lattice
temperature. In the case of 3D bulk materials, this might be
for example $N_{c/v}=2\left(m_{e/h}^{\ast}\left(T_{L}\right)k_{B}T_{n/p}/(2\pi\hbar^{2})\right)^{3/2}$,
where the effective masses are allowed to depend on the lattice temperature $T_{L}$. Moreover, the band gap energy and the band edge energies $E_{c/v}$ depend solely on
$T_{L}$. The free energy density of the phonon gas $f_{L}=f_{L}\left(T_L\right)$
and the electrostatic interaction energy $f_{\text{Coul}}=f_{\text{Coul}}\left(p-n\right)$
are the same as in Sec.~\ref{sec: Kelvin formula for the Seebeck coefficient}.

The multi-temperature free energy density $f\left(n,p,T_{n},T_{p},T_{L}\right)=f_{\text{e--h}}\left(n,p,T_{n},T_{p},T_{L}\right)+f_{\text{Coul}}\left(p-n\right)+f_{L}\left(T_{L}\right)$
is a thermodynamic potential, that yields expressions for the chemical
potentials
\begin{subequations} \label{eq: multi-temperature chemical potentials}
\begin{align}
+\mu_{c} & =\frac{\partial f}{\partial n}=k_{B}T_{n}\mathscr{F}^{-1}\left(\frac{n}{N_{c}\left(T_{n},T_{L}\right)}\right)+E_{c}\left(T_{L}\right)-q\phi \label{eq: multi-temperature chemical potentials - conduction band} \\
-\mu_{v} & =\frac{\partial f}{\partial p}=k_{B}T_{p}\mathscr{F}^{-1}\left(\frac{p}{N_{v}\left(T_{p},T_{L}\right)}\right)-E_{v}\left(T_{L}\right)+q\phi \label{eq: multi-temperature chemical potentials - valence band}
\end{align}
\end{subequations}
and the entropy densities
\begin{align*}
s_{n} & =-\frac{\partial f}{\partial T_{n}}, & s_{p} & =-\frac{\partial f}{\partial T_{p}}, & s_{L} & =-\frac{\partial f}{\partial T_{L}}.
\end{align*}
The state equations for the carrier densities follow from Eq.~\eqref{eq: multi-temperature chemical potentials} as
\begin{align*}
n & =N_{c}\left(T_{n},T_{L}\right)\mathscr{F}\left(\frac{\mu_{c}+q\phi-E_{c}\left(T_{L}\right)}{k_{B}T_{n}}\right), & p & =N_{v}\left(T_{p},T_{L}\right)\mathscr{F}\left(\frac{E_{v}\left(T_{L}\right)-q\phi-\mu_{v}}{k_{B}T_{p}}\right).
\end{align*}
In analogy to Eq.~\eqref{eq: current densities-2}, we postulate the generalized electrical current density expressions
\begin{align}
\mathbf{j}_{n} & =-\sigma_{n}\left(\nabla\varphi_{n}+P_{n,n}\nabla T_{n}+P_{L,n}\nabla T_{L}\right), & \mathbf{j}_{p} & =-\sigma_{p}\left(\nabla\varphi_{p}+P_{p,p}\nabla T_{p}+P_{L,p}\nabla T_{L}\right), \label{eq: multi-temperature current density}
\end{align}
with the thermopowers given by the Kelvin formula
\begin{align}
P_{n,n} & =-\frac{1}{q}\frac{\partial s_{n}}{\partial n}, & P_{L,n} & =-\frac{1}{q}\frac{\partial s_{L}}{\partial n}, & P_{p,p} & =+\frac{1}{q}\frac{\partial s_{p}}{\partial p}, & P_{L,p} & =+\frac{1}{q}\frac{\partial s_{L}}{\partial p}. \label{eq: multi-temperature Kelvin formulae}
\end{align}
Here, we do not describe the corresponding hot carrier transport system in full detail, but rather focus
on the central property of the electrical current densities, which is the simple representation in the drift-diffusion form (without thermodiffusion).
If the thermal driving forces can be absorbed in a generalized Einstein relation for the diffusion coefficient, this enables 
a generalized Scharfetter--Gummel discretization of type \eqref{eq: log mean temp Chatard scheme} (\emph{modified thermal voltage scheme}).
With the definitions given above, the drift-diffusion form of Eq.~\eqref{eq: multi-temperature current density} reads (along the lines in Sec.~\ref{sec: Drift-diffusion-current-densities})
\begin{align*}
\mathbf{j}_{n}&=-\sigma_{n}\left(-\frac{1}{q}\nabla\mu_{c}+P_{n,n}\nabla T_{n}+P_{L,n}\nabla T_{L}\right)=-\sigma_{n}\left(-\frac{1}{q}\nabla\frac{\partial f}{\partial n}+P_{n,n}\nabla T_{n}+P_{L,n}\nabla T_{L}\right)\\
&=-\sigma_{n}\left(-\frac{1}{q}\nabla\frac{\partial f_{\text{Coul}}}{\partial n}-\frac{1}{q}\nabla\frac{\partial f_{\text{e--h}}}{\partial n}+P_{n,n}\nabla T_{n}+P_{L,n}\nabla T_{L}\right)\\
&=-\sigma_{n}\left(\nabla\phi-\frac{1}{q}\frac{\partial^{2}f_{\text{e--h}}}{\partial n^{2}}\nabla n-\frac{1}{q}\frac{\partial^{2}f_{\text{e--h}}}{\partial n\,\partial T_{n}}\nabla T_{n}-\frac{1}{q}\frac{\partial^{2}f_{\text{e--h}}}{\partial n\,\partial T_{L}}\nabla T_{L}+P_{n,n}\nabla T_{n}+P_{L,n}\nabla T_{L}\right)\\
&=-\sigma_{n}\left(\nabla\phi-\frac{1}{q}\frac{\partial^{2}f_{\text{e--h}}}{\partial n^{2}}\nabla n+\left[P_{n,n}+\frac{1}{q}\frac{\partial s_{n}}{\partial n}\right]\nabla T_{n}+\left[P_{L,n}+\frac{1}{q}\frac{\partial s_{L}}{\partial n}\right]\nabla T_{L}\right)\\
&=-qM_{n}n\nabla\phi+M_{n}k_{B}T_{n}g\left(\frac{n}{N_{c}\left(T_{n},T_{L}\right)}\right)\nabla n
\end{align*}
(holes analogously). As a result, the generalized Einstein relation -- which now yields a nonlinear diffusion coefficient that depends on both the lattice and the carrier temperature -- allows to absorb the thermal driving forces also in the multi-temperature case, if the respective Kelvin formulae \eqref{eq: multi-temperature Kelvin formulae} for the Seebeck coefficients are assumed.

\section{Discretization of the heat source term} \label{sec: Discretization of the heat source term}

This section explains the finite volume discretization of the heat source term \eqref{eq: discrete heat generation rate} entering the discrete heat equation \eqref{eq: discrete heat equation} in more detail. First, the discrete recombination heating term is obtained by integration over the $K$-th Vorono\"i cell as
\[
H_{R,K}=\frac{1}{\vert\Omega_{K}\vert}\int_{\Omega_{K}}\mathrm{d}V\,q\left(\varphi_{p}-\varphi_{n}+\Pi_{p}-\Pi_{n}\right)R\approx q\left(\varphi_{p,K}-\varphi_{n,K}+T_{K}\left(P_{p,K}-P_{n,K}\right)\right)R_{K},
\]
using the local values of the quasi-Fermi potentials, temperature, Seebeck coefficients and $R_{K}=R(\phi_{K},\varphi_{n,K},\varphi_{p,K},T_{K})$.

The discretization of the Joule and Thomson--Peltier heating terms is more involved. The continuous expressions
\begin{align*}
H_{J} & =\sum_{\lambda\in\left\{ n,p\right\} }\frac{1}{\sigma_{\lambda}}\left\Vert \mathbf{j}_{\lambda}\right\Vert ^{2}=-\sum_{\lambda\in\left\{ n,p\right\} }\mathbf{j}_{\lambda}\cdot\left(\nabla\varphi_{\lambda}+P_{\lambda}\nabla T\right), & H_{\text{T--P}} & =-T\sum_{\lambda\in\left\{ n,p\right\} }\mathbf{j}_{\lambda}\cdot\nabla P_{\lambda}
\end{align*}
are all of the type $\tilde{H} = \kappa\left(\mathbf{j}_{\lambda}\cdot\nabla\psi\right)$,
where $\mathbf{j}_{\lambda}$ is an electrical current density vector
and $\nabla\psi$ is the gradient of a particular scalar field with $\psi\in\left\{ \varphi_{\lambda},T,P_{\lambda}\right\} $
and $\kappa\in\left\{ 1,P_{\lambda},T\right\}$ for $\lambda\in\left\{ n,p\right\}$.
The essential problem in the discretization of $\tilde{H}$ is that the full vector field $\mathbf{j}_{\lambda}$ and the full gradient $\nabla\psi$ are required on the cell $\Omega_{K}$, whereas naturally only approximations of the normal components are available.
There are several methods to tackle this problem, see Ref. \cite{Bradji2008} for a discussion. Here, we follow the approch suggested by Eymard \& Gallou\"{e}t \cite{Eymard2003}, who introduced a weakly converging gradient.
The discrete expressions \eqref{eq: discrete Joule heating} and \eqref{eq: discrete Thomson Peltier heating} are obtained along the following lines:

We seek for a finite volume discretization of the term $\tilde{H}=\kappa\left(\mathbf{j}_{\lambda}\cdot\nabla\psi\right)$ by integration over the $K$-th cell
\[
\int_{\Omega_{K}}\mathrm{d}V\,\tilde{H}=\frac{1}{2}\sum_{L\in\mathcal{N}\left(K\right)}\int_{D_{K,L}}\mathrm{d}V\,\tilde{H}\approx\frac{1}{2}\sum_{L\in\mathcal{N}\left(K\right)}\vert D_{K,L}\vert\,\kappa_{K,L}\left(\mathbf{j}_{\lambda}\cdot\nabla\psi\right)\vert_{K,L},
\]
where the integral was recast in an integral
over the adjacent bi-hyperpyramids $D_{K,L}$ (see Fig.~\ref{fig: Voronoii}) with
volumes
\begin{equation}
\vert D_{K,L}\vert=\frac{1}{d}\left\Vert \mathbf{r}_{L}-\mathbf{r}_{K}\right\Vert \vert\partial\Omega_{K}\cap\partial\Omega_{L}\vert=\frac{1}{d}\left\Vert \mathbf{r}_{L}-\mathbf{r}_{K}\right\Vert ^{2}s_{K,L}.
\label{eq: bi-hyperpyramid volume}
\end{equation}
Here, $d$ is the dimensionality of the computational domain and $s_{K,L}$
is the edge factor \eqref{eq: edge-factor}. Moreover, $\kappa_{K,L}$
is a suitably chosen average of $\kappa$ on the domain $D_{K,L}$
(see below). Following Eymard \& Gallou\"{e}t \cite[Definition 2]{Eymard2003}, we approximate $\nabla \psi$ by the weakly converging discrete gradient
\begin{equation}
\nabla\psi \vert_{K,L} \approx d \frac{\psi_{L}-\psi_{K}}{\left\Vert \mathbf{r}_{L}-\mathbf{r}_{K}\right\Vert }\, \mathbf{n}_{K,L} \label{eq: weak gradient}
\end{equation}
where $\mathbf{n}_{K,L}=\left(\mathbf{r}_{L}-\mathbf{r}_{K}\right)/\left\Vert \mathbf{r}_{L}-\mathbf{r}_{K}\right\Vert$
is the (outward-oriented) normal vector and $d\in\left\{ 1,2,3\right\}$ is the 
dimensionality of the domain.
The technique has been adopted for the discretization of Joule heating in similar models
\cite{Bradji2008, Liero2015,Fuhrmann2017} and for the computation of contact currents \cite{Farrell2017}.
Finally, we combine Eqs.~\eqref{eq: bi-hyperpyramid volume}, \eqref{eq: weak gradient} and the normal projection of the current $J_{\lambda,K,L}=\left(\mathbf{r}_{L}-\mathbf{r}_{K}\right)\cdot\mathbf{j}_{\lambda}$ (see Eq.~\eqref{eq: discrete current normal projection}, where $J_{\lambda,K,L}$ is given by a Scharfetter--Gummel type discretization scheme) and obtain
\[
\int_{\Omega_{K}}\mathrm{d}V\,\tilde{H} \approx\frac{1}{2}\sum_{L\in\mathcal{N}\left(K\right)}s_{K,L}\,\kappa_{K,L}\,J_{\lambda,K,L}\left(\psi_{L}-\psi_{K}\right),
\]
cf. Eqs.~\eqref{eq: discrete Joule heating}--\eqref{eq: discrete Thomson Peltier heating}. In order to ensure the non-negativity of the discrete Joule heating and dissipation rate (see Sec.~\ref{sec: Non-negativity-of-the discrete dissipation rate}), $\kappa_{K,L}$ is taken as the logarithmic mean temperature \eqref{eq: logarithmic mean temperature} or the edge-averaged Seebeck coefficient \eqref{eq: edge averaged Seebeck coefficients}, respectively.

\section{Power balance} \label{sec: power balance}

The balance equation for the total power is derived by integrating
the heat transport Eq.~(\ref{eq: heat equation}) over the domain
$\Omega$:
\[
\int_{\Omega}\mathrm{d}V\,c_{V}\partial_{t}T-\int_{\Omega}\mathrm{d}V\,\nabla\cdot\kappa\nabla T=\int_{\Omega}\mathrm{d}V\,H.
\]
Using Eqs.~(\ref{eq: electron transport equation})--(\ref{eq: hole transport equation})
and partial integration, the expression for the heat generation rate
(\ref{eq: heat source final}) is rearranged as
\begin{align*}
H & =-\nabla\cdot\left(\left(\varphi_{n}+\Pi_{n}\right)\mathbf{j}_{n}+\left(\varphi_{p}+\Pi_{p}\right)\mathbf{j}_{p}\right)+q\left(\varphi_{n}+\Pi_{n}\right)\partial_{t}n-q\left(\varphi_{p}+\Pi_{p}\right)\partial_{t}p.
\end{align*}
The internal energy density of the system reads $u\left(n,p,T\right)=f\left(n,p,T\right)+Ts\left(n,p,T\right)$,
where the free energy density is given in Eq.~(\ref{eq: free energy density-1}).
Using the defining relations for the quasi-Fermi potentials and the
entropy density (\ref{eq: conjugate fields-1}) as well as the Kelvin
formula (\ref{eq: Kelvin formula}), the differential internal energy
per carrier is obtained as as
\begin{align*}
\frac{\partial u}{\partial n} & =\frac{\partial f}{\partial n}+T\frac{\partial s}{\partial n}=-q\left(\varphi_{n}+\Pi_{n}\right), & \frac{\partial u}{\partial p} & =\frac{\partial f}{\partial p}+T\frac{\partial s}{\partial p}=+q\left(\varphi_{n}+\Pi_{n}\right),
\end{align*}
where the Peltier coefficients $\Pi_{n/p}$ are defined via the Kelvin
relation $\Pi_{n/p}=TP_{n/p}$. Using the definition of the heat capacity
$\heatCapacity=\partial_{T}u$, the (integrated) heat transport equation
is rearranged as a balance equation for the internal energy:
\begin{equation}
\int_{\Omega}\mathrm{d}V\,\left(\frac{\partial u}{\partial T}\frac{\partial T}{\partial t}+\frac{\partial u}{\partial n}\frac{\partial n}{\partial t}+\frac{\partial u}{\partial p}\frac{\partial p}{\partial t}\right)=\frac{\mathrm{d}}{\mathrm{d}t}\int_{\Omega}\mathrm{d}V\,u=-\oint_{\partial\Omega}\mathrm{d}A\,\mathbf{n}\cdot\left(-\kappa\nabla T+\left(\varphi_{n}+\Pi_{n}\right)\mathbf{j}_{n}+\left(\varphi_{p}+\Pi_{p}\right)\mathbf{j}_{p}\right).\label{eq: power balance prelim}
\end{equation}
In the following, we will recast the surface integral on the right
hand side into an expression for the injected electrical power and
the dissipated heat.

The local continuity equation for the internal energy (first law of
thermodynamics, cf. Eq.~(\ref{eq: power balance prelim})) reads
\begin{align}
\partial_{t}u+\nabla\cdot\mathbf{j}_{u} & =0, & \mathbf{j}_{u} & =-\kappa\nabla T+\left(\varphi_{n}+\Pi_{n}\right)\mathbf{j}_{n}+\left(\varphi_{p}+\Pi_{p}\right)\mathbf{j}_{p},\label{eq: energy conservation law}
\end{align}
where $\mathbf{j}_{u}$ is the internal energy flux density. The corresponding
continuity equation for the entropy density is obtained from Gibb's
fundamental thermodynamic relation $\mathrm{d}u=T\mathrm{d}s+\mu_{c}\mathrm{d}n-\mu_{v}\mathrm{d}p$.
By substituting Eqs.~(\ref{eq: electron transport equation})--(\ref{eq: hole transport equation})
and (\ref{eq: energy conservation law}), one arrives at
\begin{align}
\partial_{t}s & =\frac{1}{T}\partial_{t}u-\frac{\mu_{c}}{T}\partial_{t}n+\frac{\mu_{v}}{T}\partial_{t}p=-\nabla\cdot\mathbf{j}_{s}+\dot{s}_{\text{tot}}, & \mathbf{j}_{s} & =\frac{1}{T}\left(\mathbf{j}_{u}-\varphi_{n}\mathbf{j}_{n}-\varphi_{p}\mathbf{j}_{p}\right),\label{eq: entropy balance-1}
\end{align}
where $\mathbf{j}_{s}$ is the the entropy flux density. The entropy
production rate reads (cf. Sec.~\ref{sec: Non-negativity-of-the discrete dissipation rate})
\begin{equation}
\dot{s}_{\text{tot}}  =\frac{1}{T}\left(\mu_{c}-\mu_{v}\right)R+\frac{\kappa}{T^{2}}\left\Vert \nabla T\right\Vert ^{2}+\frac{1}{T}H_{J}.\label{eq: entropy production rate-2}
\end{equation}
Here, $H_{J}=\mathbf{j}_{n}\cdot\left(\nabla\varphi_{n}+P_{n}\nabla T\right)+\mathbf{j}_{p}\cdot\left(\nabla\varphi_{p}+P_{p}\nabla T\right)$
is the Joule heat term. The entropy flux density is closely connected
with the heat flux density
\begin{equation}
\mathbf{j}_{Q}=T\mathbf{j}_{s}=\mathbf{j}_{u}-\varphi_{n}\mathbf{j}_{n}-\varphi_{p}\mathbf{j}_{p}=-\kappa\nabla T+\Pi_{n}\mathbf{j}_{n}+\Pi_{p}\mathbf{j}_{p}.\label{eq: heat flux density-1}
\end{equation}
We return to the power balance equation (\ref{eq: power balance prelim})
and rewrite the surface integral on the right hand side as
\[
\oint_{\partial\Omega}\mathrm{d}A\,\mathbf{n}\cdot\left(-\kappa\nabla T+\left(\varphi_{n}+\Pi_{n}\right)\mathbf{j}_{n}+\left(\varphi_{p}+\Pi_{p}\right)\mathbf{j}_{p}\right)=\oint_{\partial\Omega}\mathrm{d}A\,\mathbf{n}\cdot\mathbf{j}_{Q}+\oint_{\partial\Omega}\mathrm{d}A\,\mathbf{n}\cdot\left(\varphi_{n}\mathbf{j}_{n}+\varphi_{p}\mathbf{j}_{p}\right).
\]
The first term is the heat flux density leaving the device. The second
term is evaluated for a device with mixed boundary conditions $\partial\Omega=\Gamma_{D}\cup\Gamma_{N}$,
where the segments $\Gamma_{D}=\bigcup_{i}\Gamma_{D,i}$ are the electrical
contacts with (ideal) Ohmic boundary conditions $\varphi_{n}=\varphi_{p}=\varphi_{D,i}=\text{const}.$
on $\Gamma_{D,i}$ (Dirichlet conditions). The remaining facets $\Gamma_{N}$
are semiconductor-insulator interfaces or artificial boundaries with
``no-flux'' boundary conditions $\mathbf{n}\cdot\mathbf{j}_{n/p}=0$.
One obtains
\[
\oint_{\partial\Omega}\mathrm{d}A\,\mathbf{n}\cdot\left(\varphi_{n}\mathbf{j}_{n}+\varphi_{p}\mathbf{j}_{p}\right)=\sum_{i}\varphi_{D,i}\int_{\Gamma_{D,i}}\mathrm{d}A\,\mathbf{n}\cdot\left(\mathbf{j}_{n}+\mathbf{j}_{p}\right)=\sum_{i}\varphi_{D,i}I_{i},
\]
where $I_{i}=\int_{\Gamma_{D,i}}\mathrm{d}A\,\mathbf{n}\cdot\left(\mathbf{j}_{n}+\mathbf{j}_{p}\right)$
is the electrical current flux across the $i$-th electrical contact.
For a two-terminal device with total current $I=I_{2}=-I_{1}$ (Kirchhoff's
current law) and applied voltage $U=\varphi_{D,1}-\varphi_{D,2}$,
this is (cf.~Ref.~\citep{Parrott1996})
\[
\oint_{\partial\Omega}\mathrm{d}A\,\mathbf{n}\cdot\left(\varphi_{n}\mathbf{j}_{n}+\varphi_{p}\mathbf{j}_{p}\right)=I_{1}\varphi_{D,1}+I_{2}\varphi_{D,2}=-UI.
\]
Thus, at stationary conditions, we finally obtain the power balance equation
as\begin{subequations}\label{eq: power balance final}
\begin{equation}
UI=\oint_{\partial\Omega}\mathrm{d}A\,\mathbf{n}\cdot\mathbf{j}_{Q}.\label{eq: power balance final 1}
\end{equation}
In conclusion, the injected electrical power $UI$ is equal to the
heat flux density leaving the device. With the divergence of the
heat flux density $\nabla\cdot\mathbf{j}_{Q}=H+\nabla\cdot\left(\Pi_{n}\mathbf{j}_{n}+\Pi_{p}\mathbf{j}_{p}\right)$,
this can also be written as
\begin{align}
UI & =\int_{\Omega}\mathrm{d}VH+\int_{\Gamma_{D}}\mathrm{d}A\,\mathbf{n}\cdot\left(\Pi_{n}\mathbf{j}_{n}+\Pi_{p}\mathbf{j}_{p}\right),\label{eq: power balance final 2}
\end{align}
\end{subequations}where the first term is the total heat generated
on the full domain and the second term describes the ``Peltier power'',
that can be either positive or negative, depending on the direction
of the current flow. In the case of optoelectronic devices (with open
optical cavities), the power balance must be supplemented by additional
terms describing the emitted radiation power \citep{Wenzel2017}.
The power balance equation (\ref{eq: power balance final}) is consistent
with the results previously reported in the literature, i.e., the
result persists when assuming the Kelvin formula for the Seebeck coefficient.

%\bibliographystyle{elsarticle-num-names}
%\bibliography{literature}

\end{document}